\documentclass[a4paper,fleqn,usenatbib]{mnras}

\usepackage{newtxtext,newtxmath}

\usepackage[T1]{fontenc}
\usepackage{ae,aecompl}

\usepackage{float}

\usepackage{graphicx}	
\usepackage{amsmath}	
\usepackage{amssymb}	
\usepackage{adjustbox} 
\usepackage{caption}
\usepackage{subcaption}

\newcommand{\alpham}{\alpha_\mathrm{M}}
\newcommand{\alphad}{\alpha_\mathrm{\Delta}}
\newcommand{\Me}{M$_\oplus$}
\newcommand{\Mstar}{M_\star}
\newcommand{\aplt}{a_\mathrm{p}}
\newcommand{\aplti}{a_\mathrm{p1}}
\newcommand{\apltii}{a_\mathrm{p2}}
\newcommand{\Mplt}{M_\mathrm{p}}
\newcommand{\Rplt}{R_\mathrm{p}}
\newcommand{\Rhill}{R_\mathrm{H}}
\newcommand{\Rhillm}{R_\mathrm{H,m}}
\newcommand{\dachaos}{\delta_\mathrm{chaos}}
\newcommand{\ap}{a_\mathrm{p}}

\newcommand{\ep}{e_\mathrm{p}}

\newcommand{\Tplt}{T_\mathrm{p}}
\newcommand{\Ti}{T_\mathrm{p1}}

\newcommand{\fin}{f_\mathrm{in}}

\newcommand{\fej}{f_\mathrm{ej}}
\newcommand{\tej}{t_\mathrm{ej}}

\newcommand{\faccplt}{f_\mathrm{acc,p}}
\newcommand{\facc}{f_\mathrm{acc}}
\newcommand{\stenau}{\Sigma(10~\mathrm{AU})}

\title[Scattering of exocomets by a planet chain]{Scattering of exocomets by a planet chain: exozodi levels and
  the delivery of cometary material to inner planets}

\author[]{Sebastian Marino$^{1}$\thanks{E-mail: s.marino@ast.cam.ac.uk}, Amy Bonsor$^{1}$, Mark C. Wyatt$^{1}$ and Quentin Kral$^{1}$  \\
  $^{1}$Institute of Astronomy, University of Cambridge, Madingley Road, Cambridge CB3 0HA, UK\\
}

\begin{document}
\label{firstpage}
\pagerange{\pageref{firstpage}--\pageref{lastpage}}
\maketitle

\begin{abstract}

Exocomets scattered by planets have been invoked to explain
observations in multiple contexts, including the frequently found
near- and mid-infrared excess around nearby stars arising from
exozodiacal dust. Here we investigate how the process of inward
scattering of comets originating in an outer belt, is affected by the
architecture of a planetary system, to determine whether this could
lead to observable exozodi levels or deliver volatiles to inner
planets. Using N-body simulations, we model systems with different
planet mass and orbital spacing distributions in the 1-50~AU
region. We find that tightly packed ($\Delta \ap<20 \Rhillm$) low mass
planets are the most efficient at delivering material to exozodi
regions (5-7\% of scattered exocomets end up within 0.5~AU at some
point), although the exozodi levels do not vary by more than a factor
of $\sim7$ for the architectures studied here. We suggest that
emission from scattered dusty material in between the planets could
provide a potential test for this delivery mechanism. We show that the
surface density of scattered material can vary by two orders of
magnitude (being highest for systems of low mass planets with medium
spacing), whilst the exozodi delivery rate stays roughly constant, and
that future instruments such as JWST could detect it. In fact for
$\eta$~Corvi, the current Herschel upper limit rules out the
scattering scenario by a chain of $\lesssim$30~\Me\ planets. Finally,
we show that exocomets could be efficient at delivering cometary
material to inner planets (0.1-1\% of scattered comets are accreted
per inner planet). Overall, the best systems at delivering comets to
inner planets are the ones that have low mass outer planets and medium
spacing ($\sim20\Rhillm$).

\end{abstract}




\section{Introduction}
\label{sec:intro}


Planetesimal belts have been inferred in multiple systems to explain
the now common infrared (IR) excess found around main-sequence stars,
which originates from circumstellar dust \citep[e.g.,][]{Su2006,
  Hillenbrand2008, Carpenter2009, Eiroa2013, Absil2013,
  Matthews2014pp6, Thureau2014, Montesinos2016}. Mutual collisions
within these Asteroid or Kuiper belt analogues grind down solids
sustaining high levels of dust on timescales as long as Gyr's, giving
rise to debris discs \citep[e.g.,][]{Dominik2003,
  Wyatt2007collisionalcascade}.


While most of the systems with cold and warm dust are consistent with
Asteroid or Kuiper belt analogues located beyond a few AU
\citep[e.g.,][]{Wyatt2007collisionalcascade, Schuppler2016,
  Geiler2017}, a few systems present levels of warm/hot dust detected
at near- and mid-infrared (NIR and MIR) wavelengths (exozodis) that
are incompatible with such a scenario \citep[see review by
][]{Kral2017exozodis}. This is because there are limits on how bright
a debris disc can be at any given age
\citep[][]{Wyatt2007hotdust}. While at tens of AU a belt of km-sized
or larger planetesimals can survive against collisions on Gyr
timescales, within a few AU debris discs collisionally evolve much
faster setting a limit on their brightness. Well studied systems like
the 450 Myr old A-star Vega \citep{Absil2006, Defrere2011, Su2013} and
the 1-2 Gyr old F-star $\eta$~Corvi \citep{Stencel1991, Wyatt2005,
  Smith2009etacorvi} have exozodi levels above this limit. This
incompatibility implies that the dust in their inner regions or
material sustaining it must be either transient, e.g. as a result of a
giant collision between planetary embryos or a giant impact on a
planet \citep[e.g.,][]{Jackson2014, Kral2015}, or be continually fed
from material formed further out in the system where it can survive
for much longer timescales. In fact, this is the case for the zodiacal
cloud, much of which is replenished from dust that originates from
comets that disintegrate as they pass through the inner Solar System
\citep{Nesvorny2010}. Based on the NIR excess measured in nearby stars
using interferometry, it has been estimated that $\sim10-30\%$ of
AFGK-type stars have exozodi levels above $1\%$ of the stellar flux
\citep{Absil2013,Ertel2014}, with a tentative correlation with the
existence of an outer reservoir of cold dust for FGK-type stars
\citep{Nunez2017}. On the other hand, MIR excess are less commonly
detected \citep[e.g.][]{Kennedy2013}, but at the 1\% excess level
there again appears to be a correlation with the presence of an outer
belt \citep{Mennesson2014}.

Further observational evidence for inward transport of material in
planetary systems, but in a different astrophysical context, comes from
the presence of elements heavier than He in the atmospheres of
$\sim30\%$ of white dwarfs \citep[the so-called polluted white dwarfs,
  e.g.][]{Zuckerman2003, Zuckerman2010, Koester2014}, which suggests
that these must be accreting solid material formed beyond a few AU
where it could have survived the AGB phase \citep{Farihi2010,
  Debes2002}. Moreover, some of these polluted white dwarfs also show
infrared excess and the presence of metallic gas within their Roche
limit \citep[e.g.,][]{Gansicke2006, Melis2010}, indicating the
presence of circumstellar material accreting onto the white dwarf.

Different mechanisms have been invoked to explain the bright hot dust,
and possibly the white dwarf pollution phenomena: inward scattering of
solids from an outer debris belt by a chain of planets
\citep[e.g.,][]{Bonsor2012analytic}; planetesimals evolving into
cometary orbits due to mean motion resonances with an exterior high
mass planet on an eccentric orbit \citep{Beust1996, Faramaz2017};
instabilities in the system after which a planet could disrupt a
planetesimal belt scattering large amounts of material inwards
\citep[similar to the Late Heavy Bombardment in the Solar
  System,][]{Booth2009, Bonsor2013}. The latter has been particularly
studied to explain the infrared excess around polluted white dwarfs
\citep[e.g.,][]{Debes2002, Veras2013instability}.

Alternatively, it has also been considered that exozodis could be fed
naturally by the small dust that is continually produced through
collisions in an outer debris belt and that migrates in through P-R
drag \citep[e.g.,][]{vanLieshout2014, Kennedy2015prdrag}. Moreover,
P-R drag models predict that significant amounts of $\mu$m-sized dust
should lie within exozodis and outer belts, even in the absence of
planets, and distributed with a characteristic flat surface density
that could even hinder the characterisation of inner planets. However,
P-R drag models fail to reproduce the high exozodi levels for some
extreme systems such as $\eta$~Corvi. This is because dust undergoing
P-R drag and sublimation does not concentrate in high enough levels in
the inner regions, although this could be circumvented by magnetic
trapping \citep{Rieke2016}.


Understanding the inward transport of material in exoplanetary systems
is of great importance as it can set constraints on the architecture
of planetary systems \citep[e.g.][]{Bonsor2012analytic,
  Bonsor2012nbody}, but also it can be used to assess the possibility
that material formed at large radii, i.e. rich in volatiles in the
form of ices and organic molecules, could be delivered to inner
planets via impacts. Impacts and volatile delivery by comets and
asteroids originating in the outer asteroid belt has been proposed to
account for the water on Earth as it might have formed dry
\citep{Morbidelli2000, Obrien2006, Raymond2009}. In the particular
case of the Solar System, the formation of Jupiter and Saturn could
have been determinant in the delivery of water and volatiles to its
inner regions \citep{Raymond2017}. Moreover, impacts could have also
acted as chemical activators and the energy source for amino acid
formation in primitive atmospheres \citep[e.g.,][]{Civis2004,
  Trigo-Rodriguez2017}. For other systems with terrestrial planets
formed closer in, volatile delivery could be essential for the
development of an atmosphere that could support life
\citep[e.g.,][]{Raymond2007, Lissauer2007}. Thanks to the
unprecedented sensitivity and resolution of the Atacama Large
Millimeter/submillimeter Array (ALMA), it has been possible to detect
exocometary gas and constrain the volatile composition of
planetesimals in exo-Kuiper belts \citep[e.g.,][]{Dent2014,
  Marino2016, Kral2016, Marino2017etacorvi, Matra2017betapic,
  Matra2017fomalhaut}. This supports the general picture that
volatiles can be locked in planetesimals (similar to Solar System
comets) that could be later delivered to planets \citep{deNiem2012}.


In this paper we study how the scattering of solids
(e.g. planetesimals, dust, etc) varies as a function of the
architecture of a planetary system. In particular, we focus on
particles on unstable orbits originating from an outer belt (e.g. an
exo-Kuiper belt) that get scattered by an outer planet and can
continue being scattered to the inner regions of a system by a chain
of planets. We use N-body simulations to investigate:
\begin{enumerate}
\item whether or not exozodis can be produced by the inward
  scattering of material from an outer planetesimal belt, and how this
  ability depends on the architecture of the planetary system. There
  are two observational constraints that must be met, the amount of
  material delivered to the inner regions must be able to account for
  the observed excess, and the amount of material in between the outer
  belt and exozodi must not exceed the upper limits from observations.
\item the planetary system architectures that are best suited for
  delivering material to inner planets, including volatiles locked in
  ices and organic material.
\end{enumerate}

In \S\ref{sec:framework} we describe the general framework of the
inward transport problem, and on which aspects we will focus in this
paper.  In \S\ref{sec:analytic} we summarise the key factors and
considerations that can increase or reduce the inward transport of
material via scattering and the number of impacts on inner
planets. Then in \S\ref{sec:sim} we describe the initial conditions of
a set of N-body simulations that we use to tackle the two points
above. \S\ref{sec:results} presents the results from the
simulations. In \S\ref{sec:discussion} we discuss the observability of
material being scattered between the planets using current and future
instruments and implications for one known system, which systems are
potentially the most optimum at producing exozodis and delivering
cometary material to inner planets, and we test how a different choice
of simulation parameters could affect our results. Finally,
\S\ref{sec:conclusions} summarises our conclusions.






 \begin{figure}
  \centering
 \includegraphics[trim=0.0cm 0.0cm 0.0cm 0.0cm, clip=true, width=1.0\columnwidth]{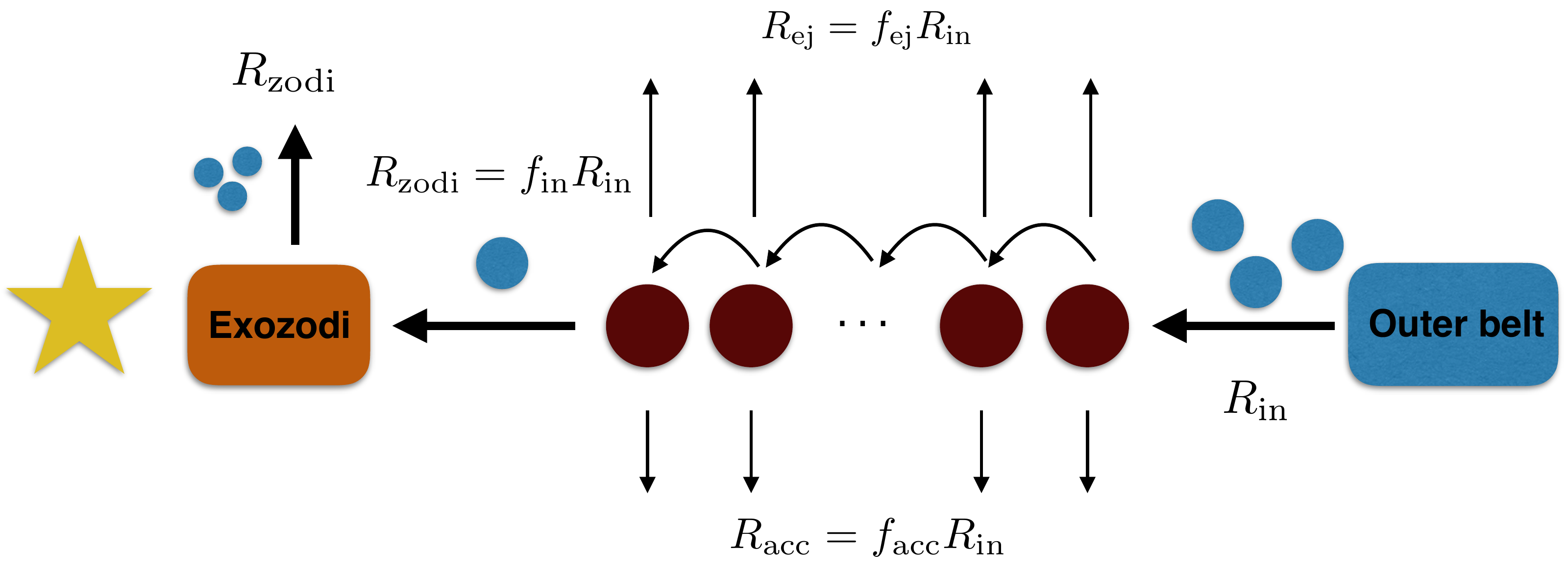}
 \caption{Sketch of a planetary system (planets as brown circles)
   scattering planetesimals (blue circles) that are input near the
   outermost planet at a rate $R_\mathrm{in}$, get ejected at a rate
   $R_\mathrm{ej}$, accreted by planets at a rate $R_\mathrm{acc}$ and
   get to the inner regions at a rate $R_\mathrm{zodi}$. Once in the
   inner regions, we assume that planetesimals are processed into
   small dust that is lost, e.g. via radiation pressure or P-R drag.}
 \label{fig:sketch}
\end{figure}

\section{Our framework}

\label{sec:framework}
The process of scattering of particles by a chain of planets that feed
an exozodi can be split into three parts (see sketch in Figure
\ref{fig:sketch}): \begin{enumerate}
\item Planetesimals are put on unstable orbits near the outermost
  planet. Planetesimals could be born there or be transported from an
  outer belt, e.g. via chaotic diffusion \citep{Morbidelli2005}. We
  simply assume a constant input rate $R_\mathrm{in}$.
\item Planetesimals are scattered by the chain of planets resulting in
  ejections, accretion onto planets, or inward transport.
\item The solid mass in planetesimals that gets to the inner regions
  where the exozodi lies, is assumed to be transformed into dust and
  removed. This could happen through sublimation of ices or disruption
  events, releasing dust as Solar System comets, or through mutual
  collisions in a so-called collisional cascade.
\end{enumerate}

In this paper we focus on studying the second stage (ii). We want to
test the analytic predictions from previous work (see Section
\ref{sec:analytic} below) and investigate further the process of
scattering by multiple planets on circular orbits. We assume that
particles are already on unstable orbits, or equivalently, they are
input near the outermost planet at a constant rate $R_\mathrm{in}$
(see possible scenarios in \S\ref{sec:disinwards}). We follow the
different outcomes of particles being scattered and we trace the
number of particles that get ejected, accreted and that reach the
exozodi region. We assume that particles are lost immediately after
reaching this exozodi region (or rather on a timescale much shorter
than their orbital evolution). This is the most optimistic scenario as
in reality only a fraction of the mass would end up as exozodiacal
dust.

\section{Scattering considerations}
\label{sec:analytic}

In this section we describe the results from previous studies that we
use to make predictions regarding how the scattering process depends
on the architecture of a planetary system. The basic condition for a
particle to be scattered by a planet is that their orbits must cross
or get sufficiently close in order to have a close encounter. For a
planet on a circular orbit this translates to a condition on the
planet's semi-major axis ($\ap$) and on the semi-major axis and
eccentricity of the particle ($a$ and $e$, respectively), i.e.
\begin{equation}
  a(1+e)\gtrsim \ap - 1.5\Rhill=Q_{\min} \ \mathrm{if}\ a<\ap, \label{eq:Qmin}
\end{equation}
or
\begin{equation}
  a(1-e)\lesssim \ap + 1.5\Rhill=q_{\max} \ \mathrm{if}\ a>\ap, \label{eq:qmax}
\end{equation}
where $\Rhill=\aplt(\Mplt/(3M_\star))^{1/3}$ is the planet's Hill
radius and $\Mplt$ is the planet mass. The factor preceding $\Rhill$
should be of the order of unity and is arbitrarily set to 1.5 to match
the results presented below. Particles that satisfy this condition can
be scattered by the planet diffusing in energy or $1/a$
\citep[e.g.,][]{Duncan1987}. Below we present the main analytic
considerations and predictions from previous works that we will test
with our simulations.

\subsection{Planet spacing - multiple scattering}
\label{sec:tiss}


Particles scattered by only one planet on a circular orbit will be
constrained by the Jacobi constant or Tisserand parameter
\citep[$\Tplt$,][]{Tisserand1896, Murray&Dermott1999}, which is
conserved in the circular restricted three-body problem when $\Mplt\ll
M_{\star}$. The Tisserand parameter can be written as
\begin{equation}
  \Tplt= \frac{\aplt}{a} + 2 \sqrt{\frac{(1-e^2)a}{\aplt}}\cos(I), \label{eq:tiss}
\end{equation}
where $I$ is the particle's inclination. Particles with
$\Tplt\lesssim3$ can get sufficiently close to a planet (e.g. within a
Hill radius) to experience a close encounter. As shown by
\cite{Bonsor2012analytic}, the condition for scattering together with
the conservation of $\Tplt$ implies that a particle being scattered by
a single planet on a circular orbit will be constrained in the $a-e$
space. Specifically, given the restriction in Equation \ref{eq:Qmin},
zero inclination and assuming $2<\Tplt<3$, there is a minimum
pericentre that a particle can reach ($q_{\min}$) given by
\begin{footnotesize}
\begin{equation}
  q_{\min}=\frac{- \aplt Q_{\min} \Tplt^2 + 2\aplt^2\Tplt + 4Q_{\min}^2-4\sqrt{ 2\aplt^3Q_{\min} - \aplt^2Q_{\min}^2\Tplt+ Q_{\min}^4}}{\aplt\Tplt^2-8Q_{\min}}. \label{eq:qmin}
\end{equation}
\end{footnotesize}
Equation \ref{eq:qmin} implies that particles initially in low
eccentricity orbits ($\ep\lesssim0.3$) cannot reach the very inner
regions if their Tisserand parameter is conserved, i.e. when being
scattered by a single planet.

There are multiple ways around this restriction. Particles could
arrive near the scattering region of the planet with high
eccentricities or inclinations, i.e. low $\Tplt$, for example if
originating in an exo-Oort cloud. Alternatively, the presence of
additional planets could modify the initial Tisserand parameter of
particles if these get scattered by multiple planets, in which case
there is a constraint on the separation of the additional planets
\citep{Bonsor2012analytic}. This is the scenario that is considered in
our modelling approach.

Assuming particles start with low eccentricity and inclinations near
the outermost planet (which is the case for our simulations), for
efficient inward scattering the next planet in the chain must have a
semi-major axis larger or near $q_\mathrm{min,p1}$, which is the
minimum pericentre that particles can reach when scattered by the
outermost planet (``p1''). If not, inward scattering may still occur
since particles could have their Tisserand parameter modified via
secular perturbations or resonances from the additional planets,
although these act on longer timescales. Therefore, assuming particles
start on orbits with low $e$ and $I$, and near planet p1 ($\Ti$ close
to 3), our first consideration for a chain of planets to optimally
scatter particles inwards from an outer belt, is that their two
outermost planets must be in orbits near to each other.  The closer
they are, the higher the number of particles with $\Ti$ small enough
to reach the next planet in the chain. If particles are initially in
highly eccentric or inclined orbits, e.g. Oort cloud objects, their
Tisserand parameter will be low enough such that the separation
between the planets is no longer a constraint.

\begin{figure}
  \centering
 \includegraphics[trim=0.5cm 0.0cm 0.5cm 0.0cm, clip=true, width=1.0\columnwidth]{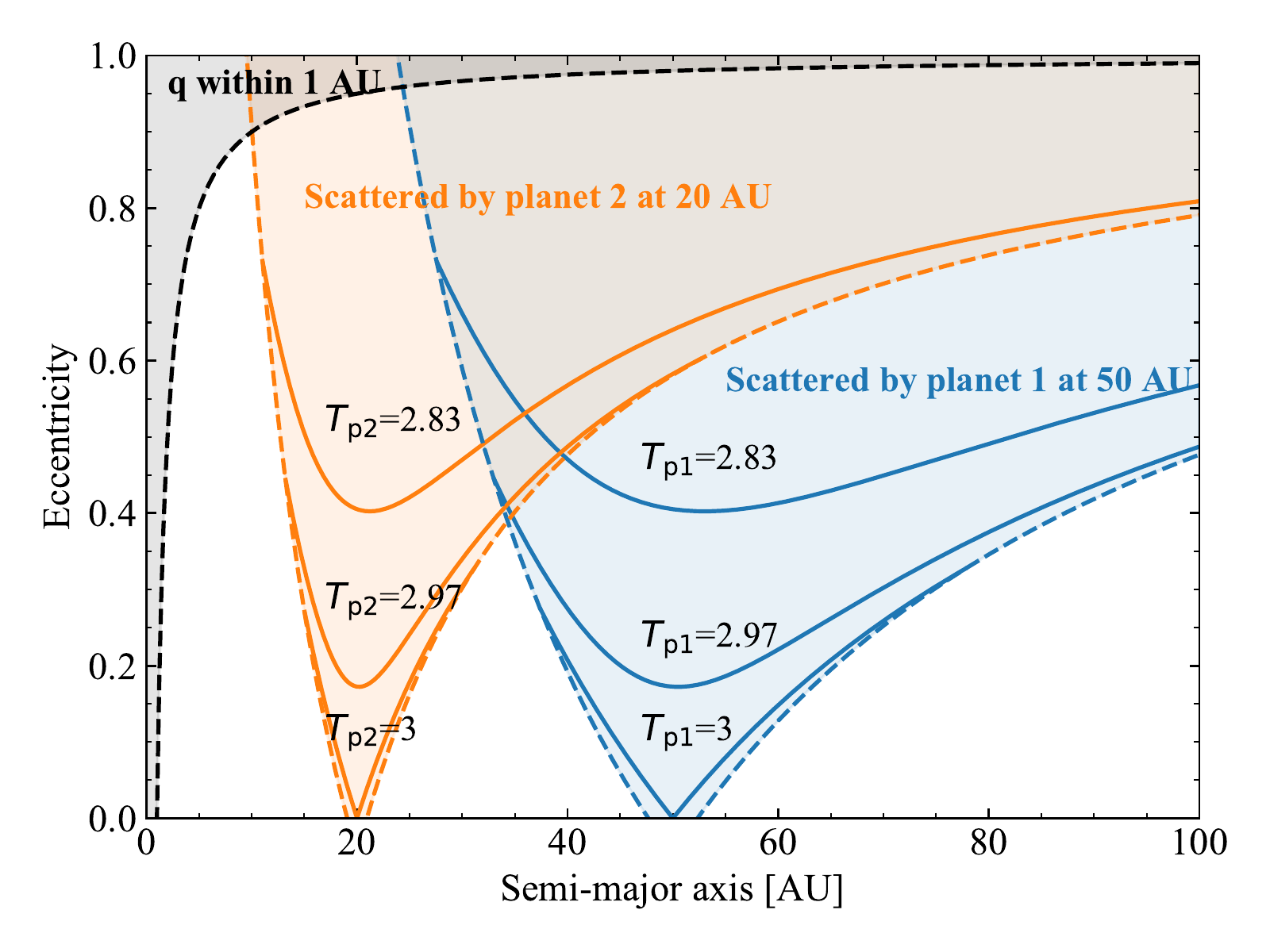}
 \caption{Constant Tisserand parameter curves and scattering regions
   for two 30 M$_\oplus$ planets in circular orbits and with
   semi-major axes of 50 (i, blue) and 20 AU (ii, orange). The dashed
   lines correspond to the scattering regions, i.e. orbits crossing
   $\aplt\pm1.5\Rhill$. The continuous lines represent curves of
   constant Tisserand parameter with respect to planets i and ii, and
   assuming $I=0$. The black dashed line shows the semi-major axes and
   eccentricities of particles that have a pericentre lower than 1 AU.}
 \label{fig:scatTiss}
\end{figure}

To illustrate the scattering restriction, Figure \ref{fig:scatTiss}
shows curves of constant Tisserand parameter (solid lines) as a
function of semi-major axis and eccentricity with respect to two 30
\Me \ planets on circular orbits with $\aplt=20,50$ AU (orange and
blue, respectively). Assuming particles in the system are initially in
the vicinity of the outermost planet (denoted as p1), their final
outcome will depend on their initial Tisserand parameter with respect
to p1. For example, $\Ti\gtrsim3.0$ particles will form a scattered
disc constrained within $\ap=35-80$ AU and since particles with
eccentricities lower than 0.4 will never have a close encounter with
the second planet (p2), these will not make it to the inner
regions. Otherwise, $\Ti\lesssim2.97$ particles can reach a pericentre
near 20 AU and be scattered by p2 (which then changes their
$\Ti$). Multiple scattering can put particles on highly eccentric
orbits that can reach the inner regions (in grey for those reaching a
radius smaller than 1~AU), or get eventually ejected from the
system. This condition is met in the Solar System, where Uranus is
close enough to Neptune such that particles in the Kuiper belt with
$T_\mathrm{Nep}<2.999$ can be scattered by Uranus after a few
encounters with Neptune \citep[e.g.,][]{Levison1997}. Additional
planets could be present closer in which could help to increase the
inward flux of scattered particles.

\subsection{Planet masses - timescales}

The second consideration relates the mass distribution of planets as a
function of semi-major axis. As stated by \cite{Wyatt2017}, particles
will be passed inwards more efficiently when interior planets start
dominating the scattering process, i.e. when the scattering timescale
is shorter for interior planets. The scattering timescale can be
approximated by the cometary diffusion time \citep{Brasser2008}
\begin{equation}
   t_\mathrm{scat}\cong M_\star^{3/2} \aplt^{3/2}\Mplt^{-2}, \label{eq:teject}
\end{equation}
where $\Mplt$ is in units of Earth masses, $\aplt$ in AU,
$t_\mathrm{scat}$ in Gyr, and $M_\star$ is the mass of the central
star in solar masses, respectively. Since these timescales increase
with orbital radius and decrease with planet mass, chains of planets with
equal mass or decreasing as a function of radius should be more
efficient at passing particles inwards.

If particles are born near the outermost planet in the chain, that
planet's mass will determine the timescale and rate at which particles
are scattered in. For example, if the scattering timescale is longer
or of the order of the age of a system, then planet scattering can
sustain an exozodi at the current age of the system relying solely on
planetesimals formed in its vicinity. Otherwise, the exozodi fed by
scattered primordial particles would only last a fraction of the age
of a system. Using N-body simulations, \cite{Bonsor2012nbody} studied
this considering chains of equal mass planets and how their masses can
affect the process of inward scattering when particles start in the
unstable region near the outermost planet. They found that while
higher mass planets can scatter material inwards at a higher rate,
lower mass planets continue to scatter material inwards on longer
timescales as they clear their orbits on a longer
timescale. Therefore, in order to sustain high exozodiacal dust levels
for 0.1-1 Gyr from planetesimals born near the outermost planet, a
massive outer belt and a chain of tightly packed low mass planets were
necessary.

Similarly, using analytical arguments \cite{Wyatt2017} considered that
in order to maximise the inward flux in a system at a given age, no
planets should lie in the \textit{ejected} region in the $\aplt-\Mplt$
space. In this region planets can eject particles on a timescale
(approximated by the cometary diffusion timescales) shorter than the
age of the system, and thus might never get to the inner regions
before being ejected, or if they do, they will be quickly
removed. However, the effect of a complex multi-planet system on this
simple argument has to be investigated.

An important caveat is that implicit in the arguments above is that
planetesimals were born near the outermost planet. The scenario that
we are considering in this paper is that particles are continually
input near the outermost planet. This means that both of the above
constraints are relaxed (i.e. the scattering timescale for the
outermost planet, and those closer in, can potentially be shorter than
the system age and still result in material reaching the inner regions
at a given age. However, for planets that are very low in mass, the
orbits of particles will still evolve on timescales much longer than
the age of the system, thus, being unable to scatter particles in at a
high rate and setting a lower limit on the mass of the outermost
planet. For example, particles scattered by a 5 \Me \ planet at 50 AU
will evolve on timescales of $\sim10$~Gyr.

To conclude, we expect that the best systems for scattering particles
inwards from an outer belt should be the ones with equal mass planets
or decreasing mass with orbital radius. If we additionally require
that the particles being scattered are born in the vicinity of the
outermost planet (which is not the case here), then that outermost
planet should also lie below the ejected region for the age of the
system.

\subsection{Planet masses and relative velocities - accretion}

The third consideration relates to the possibility of particles being
accreted by planets, possibly delivering volatiles to the inner
planets. To quantify the number of particles that will be accreted by
a planet, or the rate at which they will, it is necessary to consider
the mass and radius of the planet (which define its collisional
cross-section), the volume density of particles that the planet
encounters, and the encounter velocities. If the system is continually
fed from particles starting in the outer regions (the scenario that we
are considering here), these will then form what we will call a
\textit{scattered disc}, with a density of particles that will reach a
steady state. This density represents the amount of material that can
potentially be accreted at any given time and orbital radius. It will
depend both on the amount of material that is passed in from the
outermost planet to the inner regions, and on its lifetime before
being lost, e.g. via ejection or onto the rest of the planets in the
system. The rate at which a planet accretes can be approximated as
\begin{equation}
  R_\mathrm{acc,p}=\frac{v_\mathrm{rel} \Gamma
    \Sigma(r)}{h}, \label{eq:Racc}
\end{equation}
where $\Gamma$, $\Sigma(r)$, $v_\mathrm{rel}$ and $h$ are the
collisional cross-section of the planet, the steady state surface
density of particles (number per unit area), the relative velocity
between a planet and particles, and the scale height of the scattered
disc, respectively. The latter can be approximated by the product of
$r$ and the average inclination $<I>$ of the particles orbits. Both
$\Sigma(r)$ and $h$ are determined by the specific architecture of the
planetary system. For example, particles being scattered by low mass
planets are likely to survive for longer timescales against ejection,
increasing $\Sigma(r)$. The collisional-cross section is defined by
the mass and radius of the planet, together with the relative
velocities, as
\begin{equation}
  \Gamma=\pi\Rplt^2
  \left(1+\frac{v_\mathrm{esc}^2}{v_\mathrm{rel}^2}\right), \label{eq:crosssection}
\end{equation}
where $\Rplt$, $v_\mathrm{esc}$ and $v_\mathrm{rel}$ are the radius of
the planet, its escape velocity, and relative velocity before the
encounter, respectively. The higher $\Gamma$ is, the higher the rate
of impacts on that planet (Equation \ref{eq:Racc}). For
$v_\mathrm{esc}\gg v_\mathrm{rel}$ and $\Mplt\propto\Rplt^{3}$
(i.e. fixed density) we find $\Gamma\propto\Mplt^{4/3}$, otherwise
$\Gamma\propto\Mplt^{2/3}$. Therefore, $R_\mathrm{acc,p}$ will be
greater for more massive planets as these have greater radii and
escape velocities (i.e. greater $\Gamma$), lower relative velocities
(greater $\Gamma$ and lower $h$) and higher surface densities.

However, it is unclear how the mass distribution of planets will also
affect the density of particles (determined by the inward scattering
and particles lifetimes) and their relative velocities, with the
latter mainly defined by the distribution of eccentricities and
inclinations. We can guess that chains of more massive planets will
result in lower $\Sigma(r)$ and higher relative velocities because
particles are easily put on highly eccentric/inclined orbits. As shown
by \cite{Wyatt2017}, the most likely outcome of particles being
scattered by a planet can be understood by comparing the planet's
Keplerian velocity ($v_\mathrm{k}$) with its escape velocity
$v_\mathrm{esc}$. This is because the maximum kick that a particle can
experience after a single scattering event is of the order of
$v_\mathrm{esc}$, since to get a larger kick it would have to come so
close that it would hit the planet. Therefore, if $v_\mathrm{esc}\gg
v_\mathrm{k}$ particles are likely to be ejected in a few close
encounters, decreasing the surface density of the scattered
disc. Otherwise particles would need a large number of encounters
before being ejected, increasing the surface density of the scattered
disc and the likelihood of being accreted by a planet. Equating
$v_\mathrm{esc}$ and $v_\mathrm{k}$ we find \citep[equation 1
  in][]{Wyatt2017}
\begin{equation}
\Mplt\cong 40 M_\star^{3/2} \aplt^{-3/2} \rho_\mathrm{p}^{-1/2}, \label{eq:vesck} 
\end{equation}
where $\rho_\mathrm{p}$ is the bulk density of the planet in units of
g~cm$^{-3}$ and $\Mplt$ is in units of \Me. Planets below this mass
are likely to accrete particles if these are not lost on shorter
timescales via other means.

\subsection{Predictions}
\label{sec:predictions}
To summarise, assuming particles are input at a constant rate in the
vicinity of the outermost planet, we predict the following based on
previous studies: \begin{enumerate}
\item Systems with outer planets close to each other will be better at
  scattering particles inwards as more particles will have a low
  enough Tisserand parameter to reach the second outermost planet
  (Equation \ref{eq:qmin}).
\item Higher mass planets will scatter and eject particles on shorter
  timescales (Equation \ref{eq:teject}) and could result in inefficient
  inward scattering.
\item In order to scatter particles inwards, scattering timescales of
  the inner planets must be shorter than those further out, as shown
  by Equation \ref{eq:teject}. These can be achieved by planet chains
  of equal mass or decreasing mass with orbital radius.
\item Planets will accrete more particles if they are more massive, if
  the surface density of particles around their orbits is higher and
  if particles are on low eccentricity and low inclination orbits,
  i.e. low relative velocities (Equations \ref{eq:Racc} and
  \ref{eq:crosssection}).
\end{enumerate}

There is no analytic prediction for how the surface density will
change when varying the planet masses as it depends both on the
scattering timescale and on the inward flux of material. Moreover, the
inward flux, distribution of eccentricities and inclinations, and
accretion onto planets could vary as a function of the spacing between
the planets. N-body simulations are well suited to study these effects
and test the predictions above.






\section{N-body Simulations}
\label{sec:sim}

In order to test our predictions presented above, and quantify how the
mass distribution and orbit spacing of a chain of planets affects the
inward transport of particles being scattered, we simulate such
interactions using N-body simulations. We model the gravitational
interactions with the N-body integrator \textsc{MERCURY} 6.2
\citep{Mercury62}, using the hybrid symplectic/Bulirsch-Stoer
integration algorithm. This allows us to speed up the simulations
computing distant interactions quickly, without losing precision in
close encounters. Our systems are composed of a 1 M$_\odot$ star, a
chain of planets and $10^3$ massless particles. The simulations lasted
1 Gyr, long enough such that the majority of the particles are lost
via ejections and accretion onto the star or planets. Outputs or
snapshots of the simulations are saved every $10^{4}$ or $10^{5}$ yr,
that we estimate is shorter than the scattering diffusion timescale
for most of our simulations (see \S\ref{sec:planets} below). We
set an outer boundary of $10^3$~AU and an inner boundary of 0.5 AU,
i.e particles are removed from the simulation when their apocentre is
larger than $10^3$~AU or their pericentre is lower than 0.5 AU. The
latter is set to trace the number of particles that are able to reach
the innermost regions that we are interested in, and also because our
time-step of 30 days is not small enough to accurately integrate the
orbits within this boundary. This assumes that particles are lost as
soon as they reach within this boundary by being incorporated into the
exozodi. Each simulation is run 20 independent times with random mean
anomalies, longitudes of ascending node and pericentres, and splitting
the total number of massless particles (50 test particles for each).

\begin{table*}
  \centering
  \caption{Setups of the different N-body simulations varying $M_0$
    (planet mass at 10 AU), $\alpham$ (planet mass semi-major axis
    dependence), $K_0$ (planet spacing at 10 AU in mutual Hill radii)
    and $\alphad$ (planet spacing semi-major axis
    dependence). N$_\mathrm{p}$ is the number of planets in each
    simulated system.}
  \label{tab:setups}
  \begin{adjustbox}{max width=1.0\textwidth}

  \begin{tabular}{clccccccc} 
    \hline
    \hline
    Number & Label & Colour & $\alpham$ & $\alphad$ & Masses [\Me] & $\aplt$ [AU] & $\Delta a/\Rhillm$ & N$_\mathrm{p}$ \\ 
    \hline
    0 & single planet & black &  & & 30 & 50.0 &  & 1 \\
    1 & reference & dark blue & 0.0 & 0.0 & 30 & (1.8, 4.2, 9.6, 21.9, 50.0) & 20 & 5 \\ 
    2 & incr M  & orange & 1.0 & 0.0 & (4, 6, 10, 18, 40, 150) & (1.2, 1.9, 3.2, 5.9, 13.5, 50.0) & 20 & 6 \\ 
    3 & decr M  & green &-1.0 & 0.0 & (180, 44, 19, 10, 6) & (1.7, 6.8, 15.8, 29.9, 50.0) & 20 & 5 \\ 
    4 &  incr K & red & 0.0 & 0.3 & 30 & (1.2, 1.9, 3.2, 6.2, 14.3, 50.0) & (11, 13, 16, 20, 28) & 6 \\ 
    5 & decr K & purple & 0.0 & -0.3 & 30 & (2.4, 7.1, 15.6, 29.4, 50.0) & (25, 19, 16, 13) & 5 \\ 
    6 & high M & brown & 0.0 & 0.0 & 90 & (1.1, 3.9, 13.9, 50.0) & 20 & 4 \\ 
    7 & low M  & pink & 0.0 & 0.0 & 10 & (1.0, 1.8, 3.1, 5.4, 9.4, 16.4, 28.7, 50.0) & 20 & 8 \\ 
    8 & high K & yellow & 0.0 & 0.0 & 30 & (3.4, 13.0, 50.0) & 30 & 3 \\ 
    9 & low K & light blue & 0.0 & 0.0 & 30 & (1.1, 1.8, 2.8, 4.6, 7.4, 11.9, 19.2, 31.0, 50.0) & 12 & 9 \\ 
    10 & low M - low K & grey & 0.0 & 0.0 & 10 & (1.3, 1.9, 2.6, 3.6, 5.0, 7.0, 9.7, 13.4, 18.7, 25.9, 36.0, 50.0) & 12 & 12 \\ 
    11 & very low K & light blue & 0.0 & 0.0 & 30 & (1.1, 1.6, 2.1, 2.9, 4.0, 5.5, 7.5, 10.3, 14.1, 19.4, 26.6, 36.5, 50.0) & 8 & 13 \\ 
    \hline 
  \end{tabular}
  
  \end{adjustbox}
\end{table*}

\begin{figure*}
  \centering
 \includegraphics[width=1.0\textwidth]{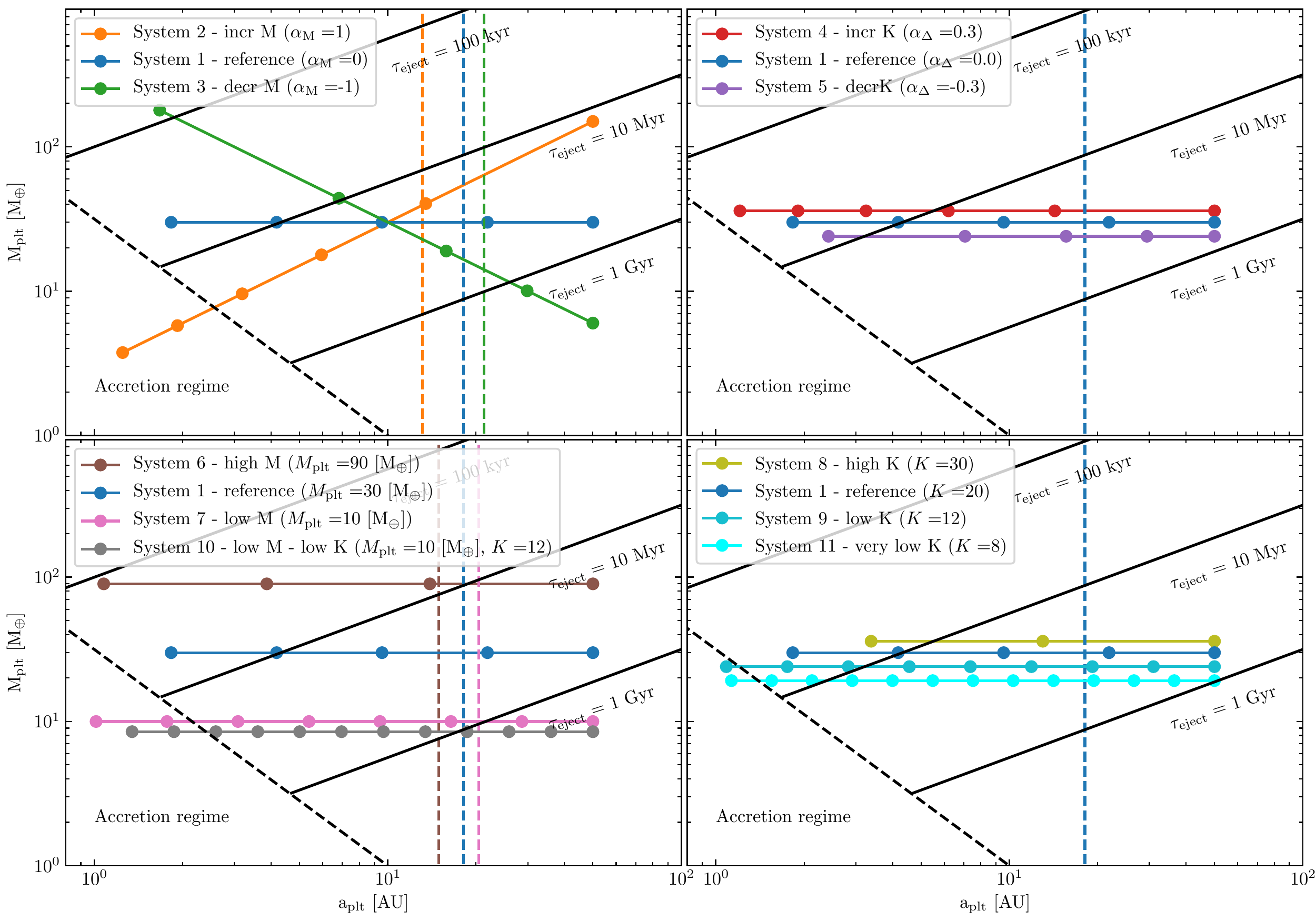}
 \caption{Masses and semi-major axes of planets in each simulated
   system. \textbf{\textit{Upper left panel}}: chains of planets with
   varying $\alpham$. \textbf{\textit{Upper right panel}}: chains of
   planets with varying $\alphad$ (the mass of the planets is offset
   for display). \textbf{\textit{Lower left panel}}: chains of equal
   mass planets of 10, 30 and 90 \Me\ and separations of 20 or 12
   mutual Hill radii. \textbf{\textit{Lower right panel}}: chains of
   equal mass planets with $K_0=$12, 20 and 30. The dots represent the
   position of the planets. The vertical dashed lines represent the
   minimum pericenter of particles scattered by the outermost planet
   given their initial Tisserand parameter (Equation
   \ref{eq:qmin}). The dashed black lines represent the mass above
   which planets are more likely to eject particles rather than
   accrete (Equation \ref{eq:vesck}). The continuous lines represent
   the planet mass above which particles are ejected on a timescale
   shorter than 1 Gyr, 10 Myr and 100 kyr (Equation
   \ref{eq:teject}). Planet masses of systems 4, 5, 8, 9 and 10 have
   been scaled by 20\% and 36\% for a better display.}
 \label{fig:setups}
\end{figure*}

\subsection{Planet mass and semi-major axis distribution}
\label{sec:planets}
We consider a chain of planets on circular co-planar orbits, with
masses varying from 1 to 200 \Me\ (0.63 M$_\mathrm{Jup}$) \ and
semi-major axes between 1 and 50 AU. We assume densities of 1.6
g~cm$^{-3}$ (Neptune's density). As we are interested in studying how
the scattering process depends on the mass distribution and spacing of
planets, we parametrized their masses as a function of $\aplt$, and
their spacings or separations as a function of $\aplt$ and their
mutual Hill radius ($\Rhillm$). More specifically, we defined
\begin{equation}
    \Mplt(\aplt)=M_0\left(\frac{\aplt}{a_0}\right)^{\alpham}, \label{eq:mplt}
\end{equation}
where $M_0$, $a_0$ and $\alpham$ define the planet masses in the
chain. The semi-major axis of the planets is defined such that
$\aplt=50$~AU for the outermost planet and the separation between
their orbits is
\begin{eqnarray}
  \Delta a &=& K(a_\mathrm{p1}, a_\mathrm{p2}) \Rhillm,\\
  K(a_\mathrm{p1}, a_\mathrm{p2}) &=& K_0 \left(\frac{a_\mathrm{p1}+a_\mathrm{p2}}{2 a_0}\right)^{\alphad}, \label{eq:k}
\end{eqnarray}
where $K_0$ and $\alphad$ control the separation of the planets and
their long-term stability. We add planets in the system from 50 to 1
AU using Equations \ref{eq:mplt} and \ref{eq:k}. The initial mean
anomaly of every planet is chosen randomly such that each simulated
system is run on 20 slightly different configurations.

Our reference chain of planets has $M_0=30$~\Me, $a_0=10$~AU,
$\alpham=0$, $\alphad=0$ and $K_0$=20. This defines a reference to
which we compare when varying the different parameters. We vary $M_0$
between 10 and 90 \Me \ to study chains of low and high mass planets,
$\alpham$ between -1 and 1 to study the effect of decreasing or
increasing planet mass as a function of $\aplt$, $\alphad=0.3$ and
-0.3 (such that no pair of planets is closer than 10 $\Rhillm$) to
study the effect of planet spacing varying with $\aplt$, and $K_0$
between 8 and 30 to study the differences between tightly packed
systems and widely spaced. The range of spacings is inspired by the
spacing distribution found for \textit{Kepler} close in multi-planet
systems, where spacings between 10-30 mutual Hill radii are the most
common \citep{Fang2013, Pu2015, Weiss2018}, although this is uncertain
for planets at large orbital radii. In addition, if $K<10$ the system
could go unstable on a timescale shorter than 1 Gyr
\citep{Chambers1996, Smith2009stability}. Table \ref{tab:setups} and
Figure \ref{fig:setups} summarises the 10 main planet configurations
that we explore. In the same figure, the dashed black lines represent
the mass above which planets are more likely to eject particles rather
than accrete (Equation \ref{eq:vesck}), while the continuous black
lines represent the scattering diffusion timescale or the planet
masses above which particles are ejected on a timescale shorter than 1
Gyr, 10 Myr and 100 kyr (Equation \ref{eq:teject}). In all the
configurations there are planets that will eject particles on
timescales shorter than or of the order of 1 Gyr, the length of our
simulations.

\subsection{Massless particles distribution}
\label{sec:particles}

Particles are initially distributed in a cold disc, with
eccentricities and inclinations uniformly distributed between $0-0.02$
and $0-10\degr$, respectively. \footnote{Note that this choice is
  arbitrary and its effect is discussed in
  appendix~\ref{sec:vareandi}}. As we are interested only in those
particles that can be scattered, we initialised all of them in the
outermost planet's chaotic zone, i.e. the unstable region of
semi-major axes surrounding the planet in which mean motion resonances
overlap. The size of the chaotic zone has been analytically estimated
to be \citep{Wisdom1980}
\begin{equation}
\dachaos=1.3\ \aplt\ \left(\frac{\Mplt}{\Mstar} \right)^{2/7}.  
\end{equation}
Within this zone, eccentricities are excited and initially
non-crossing orbits start to cross the planet's orbit, or get
sufficiently close to have close encounters and be scattered by the
planet. We initialised all the particles with semi-major axes ($\ap$)
uniformly distributed within $a_\mathrm{plt}^{i}\pm\dachaos$
($\sim45-55$ AU for a 30 \Me \ planet at 50~AU). This ensures that
particles will be dynamically excited and likely to be scattered early
on during the simulation.

The initial distribution of orbital parameters sets the range of
Tisserand parameters that the particles have with respect to the
outermost planet (p1). For a 30 \Me \ planet, $T_\mathrm{p1}$ is
initially distributed between 2.97 and 3.01 (in
\S\ref{sec:simparameters} we discuss the effect of varying the
initial $e$ and $i$). The minimum $T_\mathrm{p1}$ is the same for all
our simulations as it is set approximately by the particles with
semi-major axis equal to that of the planet (if $e$ and $I$ are
small). On the other hand, the maximum $T_\mathrm{p1}$ is set by
particles at the inner and outer edge of the chaotic zone whose size
increases with planet mass, therefore the maximum $T_\mathrm{p1}$ is
higher for more massive planets and scales approximately as
$3+5(\Mplt/\Mstar)^{4/7}/4$. As explained in \S\ref{sec:analytic},
the minimum Tisserand parameter sets the minimum pericentre
($q_\mathrm{min,p1}$) that scattered particles can have after multiple
scattering events by the outermost planet, and thus, if they can be
readily scattered by the second outermost planet. The latter enables
the further inward scattering that we are interested in.

In Figure \ref{fig:setups} we show $q_\mathrm{min,p1}$ (vertical
dashed lines) as a reference to identify those configurations that we
expect not to be optimal for inward scattering, i.e. those systems in
which the second outermost planet has semi-major axis lower than the
minimum pericentre of particles when interacting only with the
outermost planet. This pericentre is calculated using Equation
\ref{eq:qmin} and setting $\aplt=a_\mathrm{p1}$,
$Q_{\min}=a_\mathrm{p1}-1.5\Rhill$ and
$T_\mathrm{p}=\min(T_\mathrm{p1})$, where the minimum Tisserand
parameter is approximately 2.97. For low planet masses and
$T_\mathrm{p1}=2.97$ we find $q_\mathrm{min,p1}\approx a_\mathrm{p1}/2
- 9\Rhill/2$, thus lower for higher planet masses. We anticipate that
planet configurations 4 and 8 from Table \ref{tab:setups} (red and
light green in Figure \ref{fig:setups}) will be very inefficient at
inward scattering because their outermost planets are too separated
under this criterion.

While our simulations assume that particles start near the outermost
planet on low eccentricity and inclination orbits, the exact
distribution of eccentricities and inclinations would depend on the
specific mechanism that is inputting particles near the outermost
planet. Hence, a caveat in what follows is that we are assuming that
the particles' initial conditions described above are a good
approximation to their orbits at the start of the scattering
process. The specific mechanism inputting particles and initial
conditions are discussed in \S\ref{sec:disinwards} and
\ref{sec:simparameters}.




\begin{table*}
  \centering
  \caption{Results for each of the different planetary
    configurations. All the fractions are with respect to the total
    number of particles lost after 1 Gyr of simulation. The fraction
    of trojans is computed with respect to the total number of
    particles (1000). The uncertainties are estimated based on 68\%
    confidence intervals and assuming a Poisson distribution. The
    upper limits correspond to $2\sigma$ (95\% confidence upper
    limit).}
  \label{tab:results}
  \begin{adjustbox}{max width=1.0\textwidth}

  \begin{tabular}{clccccccccc} 
    \hline
    \hline
    Number & Label & Colour & $\fej$ (\%) &  $\fin$ (\%) & $\facc$ (\%) & $\faccplt$ (\%) & Fraction accreted per planet (\%) & Fraction remaining & Absolute fraction  & $\tej$ [Myr] \\
           &       &        &             &              &              &                 &                                   & after 1 Gyr (\%)   & of Trojans (\%)    &            \\ 
    \hline
    0 & single planet & black & 85.0$^{+4.0}_{-3.9}$ & <0.65 & 15.0$^{+1.8}_{-1.6}$ & - & (15.0) & 17.0$^{+1.7}_{-1.6}$ & 31.1$^{+1.9}_{-1.8}$ & 383$\pm$21.9 \\
    1 & reference & blue & 92.7$^{+3.9}_{-3.7}$ & 3.7$^{+0.9}_{-0.7}$ & 3.6$^{+0.9}_{-0.7}$ & 0.35$^{+0.20}_{-0.13}$ & (0.3, 0.4, 0.3, 0.1, 2.4) & 4.8$^{+1.0}_{-0.8}$ & 29.5$^{+1.8}_{-1.7}$ & 120$\pm$7.1 \\
    2 & incr M & orange & 95.1$^{+3.7}_{-3.5}$ & 2.9$^{+0.8}_{-0.6}$ & 2.0$^{+0.7}_{-0.5}$ & <0.15 & (0.0, 0.0, 0.0, 0.0, 0.1, 1.8) & 0.4$^{+0.4}_{-0.2}$ & 23.8$^{+1.6}_{-1.5}$ & 12$\pm$0.8 \\
    3 & decr M & green & 92.3$^{+4.4}_{-4.2}$ & 3.1$^{+1.0}_{-0.8}$ & 4.6$^{+1.2}_{-0.9}$ & 1.06$^{+0.44}_{-0.31}$ & (0.8, 1.3, 1.2, 0.4, 1.0) & 27.7$^{+2.1}_{-2.0}$ & 28.2$^{+1.8}_{-1.7}$ & 350$\pm$29.3 \\
    4 & incr K & red & 89.0$^{+4.0}_{-3.8}$ & 2.3$^{+0.8}_{-0.6}$ & 8.7$^{+1.4}_{-1.2}$ & 0.04$^{+0.12}_{-0.03}$ & (0.0, 0.0, 0.2, 0.0, 0.3, 8.3) & 11.2$^{+1.4}_{-1.3}$ & 30.4$^{+1.8}_{-1.7}$ & 340$\pm$16.8 \\
    5 & decr K & purple & 94.7$^{+3.9}_{-3.7}$ & 3.1$^{+0.8}_{-0.7}$ & 2.2$^{+0.7}_{-0.6}$ & 0.29$^{+0.25}_{-0.14}$ & (0.3, 0.3, 0.3, 0.3, 1.0) & 1.4$^{+0.6}_{-0.4}$ & 30.5$^{+1.9}_{-1.7}$ & 79$\pm$5.8 \\
    6 & high M & brown & 94.8$^{+3.7}_{-3.6}$ & 1.9$^{+0.7}_{-0.5}$ & 3.3$^{+0.8}_{-0.7}$ & 0.07$^{+0.21}_{-0.06}$ & (0.1, 0.0, 0.1, 3.0) & 0.4$^{+0.4}_{-0.2}$ & 26.6$^{+1.7}_{-1.6}$ & 28$\pm$1.3 \\
    7 & low M & pink & 89.7$^{+4.5}_{-4.3}$ & 4.5$^{+1.2}_{-1.0}$ & 5.8$^{+1.3}_{-1.1}$ & 0.45$^{+0.19}_{-0.13}$ & (0.4, 0.4, 0.4, 0.8, 0.2, 0.6, 0.8, 2.1) & 30.4$^{+2.2}_{-2.1}$ & 30.5$^{+1.9}_{-1.7}$ & 583$\pm$36.2 \\
    8 & high K & yellow & 86.9$^{+4.0}_{-3.8}$ & 1.4$^{+0.7}_{-0.5}$ & 11.7$^{+1.6}_{-1.4}$ & <0.65 & (0.0, 0.0, 11.7) & 15.2$^{+1.6}_{-1.5}$ & 30.4$^{+1.8}_{-1.7}$ & 404$\pm$22.0 \\
    9 & low K & light blue & 93.7$^{+3.7}_{-3.6}$ & 4.8$^{+1.0}_{-0.8}$ & 1.5$^{+0.6}_{-0.4}$ & 0.19$^{+0.11}_{-0.07}$ & (0.3, 0.4, 0.1, 0.0, 0.1, 0.1, 0.4, 0.0, 0.0) & 1.5$^{+0.6}_{-0.4}$ & 26.2$^{+1.7}_{-1.6}$ & 62$\pm$4.2 \\
    10 & low M - low K & grey & 90.6$^{+4.1}_{-3.9}$ & 7.3$^{+1.3}_{-1.1}$ & 2.0$^{+0.8}_{-0.6}$ & 0.12$^{+0.09}_{-0.05}$ & (0.0, 0.0, 0.2, 0.0, 0.3, 0.2, 0.2, 0.2, 0.0, 0.2, 0.3, 0.5) & 14.4$^{+1.6}_{-1.4}$ & 31.4$^{+1.9}_{-1.8}$ & 284$\pm$19.4 \\
    11 & very low K & light blue & 91.9$^{+3.7}_{-3.6}$ & 6.7$^{+1.1}_{-1.0}$ & 1.4$^{+0.6}_{-0.4}$ & 0.02$^{+0.06}_{-0.02}$ & (0.0, 0.0, 0.0, 0.0, 0.1, 0.0, 0.0, 0.0, 0.1, 0.1, 0.1, 0.4, 0.4) & 0.4$^{+0.5}_{-0.2}$ & 28.1$^{+1.8}_{-1.7}$ & 40$\pm$2.5 \\

    \hline 
  \end{tabular}
  
  \end{adjustbox}
\end{table*}

\subsection{Analysis of simulations}
\label{sec:analysis}

For each simulation, we first remove those particles that were
initially in stable tadpole and horseshoe orbits (hereafter called
Trojans, $N_\mathrm{Troj}$). We identify Trojan particles as the ones
that after 1~Myr of evolution lie within 1.2 Hill radii from the
outermost planet and with an eccentricity lower than 0.03 (typically
$20-30\%$ of simulated particles). Then we follow the evolution of the
rest of the particles (a total number $N_\mathrm{tot}$) and trace the
number of particles ejected ($N_\mathrm{ej}$), that cross the inner
boundary at 0.5 AU ($N_\mathrm{in}$), and that were accreted by
planets ($N_\mathrm{acc}$). We compare these between the different
simulations by dividing by the total number of particles that are lost
during 1 Gyr of evolution
($N_\mathrm{lost}=N_\mathrm{ej}+N_\mathrm{in}+N_\mathrm{acc}$). We
divide by $N_\mathrm{lost}$ rather than by $N_\mathrm{tot}$ (the total
number of simulated particles) because the timescales at which
particles evolve in some of the simulated systems are comparable to 1
Gyr, hence a significant number of particles have not been lost by the
end of the simulation. These particles (a total of
$N_\mathrm{tot}-N_\mathrm{lost}$) remain in the system in highly
eccentric and inclined orbits (i.e. in the scattered disc). Hence, we
compare the fraction of particles ejected
($f_\mathrm{ej}=N_\mathrm{ej}/N_\mathrm{lost}$), that cross the inner
boundary ($f_\mathrm{in}=N_\mathrm{in}/N_\mathrm{lost}$) and that are
accreted by planets ($\facc=N_\mathrm{acc}/N_\mathrm{lost}$). We also
compute the fraction of particles accreted per inner planet
($\faccplt$), defined as those with $\aplt<10$~AU. Comparing these
relative fractions is equivalent to extrapolating to $t=\infty$
assuming that the remaining particles will be lost via ejection,
accretion, or crossing the inner edge in the same proportions as the
particles that are lost by 1 Gyr, i.e. this is equivalent to assuming
that $\fej$, $\fin$, $\facc$ stay constant over time. This is not
necessarily the case as most of the particles remaining are typically
in an outer scattered disc with large semi-major axes, where they
could be more likely to be ejected than when they started near the
outermost planet. However, this helps to give an idea of the absolute
fractions if simulations were run for longer and is shown to be
representative in one of the simulations (the one with lowest mass
planets) that was run for 5 Gyr in \S\ref{sec:simparameters}. We also
report uncertainties on $\fej$, $\fin$, $\facc$, etc. based on 68\%
Poisson confidence intervals (16th and 84th percentiles) using
analytic approximations from \cite{Gehrels1986}.

In order to quantify how fast particles are lost we also compute the
half-life of ejected particles, $\tej$. This is defined as the time it
takes to eject half of the total number of particles expected to be
ejected by $t=\infty$, assuming $N_\mathrm{ej}$ will tend to
$f_\mathrm{ej}N_\mathrm{tot}$ by the end of the simulation
(i.e. assuming that $\fej$ is constant over time).




We are also interested in the spatial distribution or surface density
distribution of scattered particles within the chain of planets, i.e
from 1 to 50 AU. Because the distribution of particles evolves with
time (see \S\ref{sec:res_single} and \ref{sec:res_reference}) and on
different timescales for each planet configuration, we focus on the
steady state surface density of scattered particles, $\Sigma(r)$. This
assumes that particles are input in the system at a constant rate
inside the chaotic zone of the outermost planet. In order to use the
simulation results to mimic a steady-state input scenario, we take
each particle and randomise its initial epoch (originally t=0) to a
value between 0 and 1Gyr. Once a Gyr is reached (the initial
integration time), we loop the particle's later evolution to $t=0$. In
order to minimise random effects caused by the finite number of
particles used in these simulations ($\sim700$ excluding Trojans), we
effectively mimic each particle 200 times by randomising in terms of
mean anomaly at each time step. This leads to an effective
$\sim$140,000 particles used to calculate the surface density
distribution at each epoch.  We then average the surface density over
time to obtain the average steady state surface density distribution
$\Sigma(r)$. We estimate the uncertainty on $\Sigma(r)$ by computing
the variance of $\Sigma(r, t)$ averaged over 100 Myr time bins
(i.e. from 10 data points at each orbital radius). Finally, because
$\Sigma(r)$ is proportional to the mass input rate (the same for all
our simulations), in our analysis we compare $\Sigma(r)$ divided by
the mass input rate.  Then the surface density can be obtained by
multiplying by any mass input rate. Below, we will also use the
surface density at 10~AU, $\stenau$, as a metric to compare different
systems.

\section{Results}
\label{sec:results}

In this section we present the main results from each simulation. We
first describe the results of a case with a single 30~\Me\ planet at
50~AU (\S\ref{sec:res_single}) and our reference chain of equal
mass planets (\S\ref{sec:res_reference}). Then, we present results
for planet configurations of equal planet mass, but varying their
spacing (\S\ref{sec:res_spacing}), and configurations with constant
spacing in mutual Hill radii, but varying their masses
(\S\ref{sec:res_mass}). In \S\ref{sec:predictionscorrect} we
discuss our results in the context of the predictions made in Sec
\ref{sec:analytic}.


\begin{figure}
  \begin{tabular}[t]{|c|}
    \begin{subfigure}{1.0\columnwidth}
      \includegraphics[trim=0.0cm 0.0cm 0.0cm 0.0cm, clip=true, width=1.0\columnwidth]{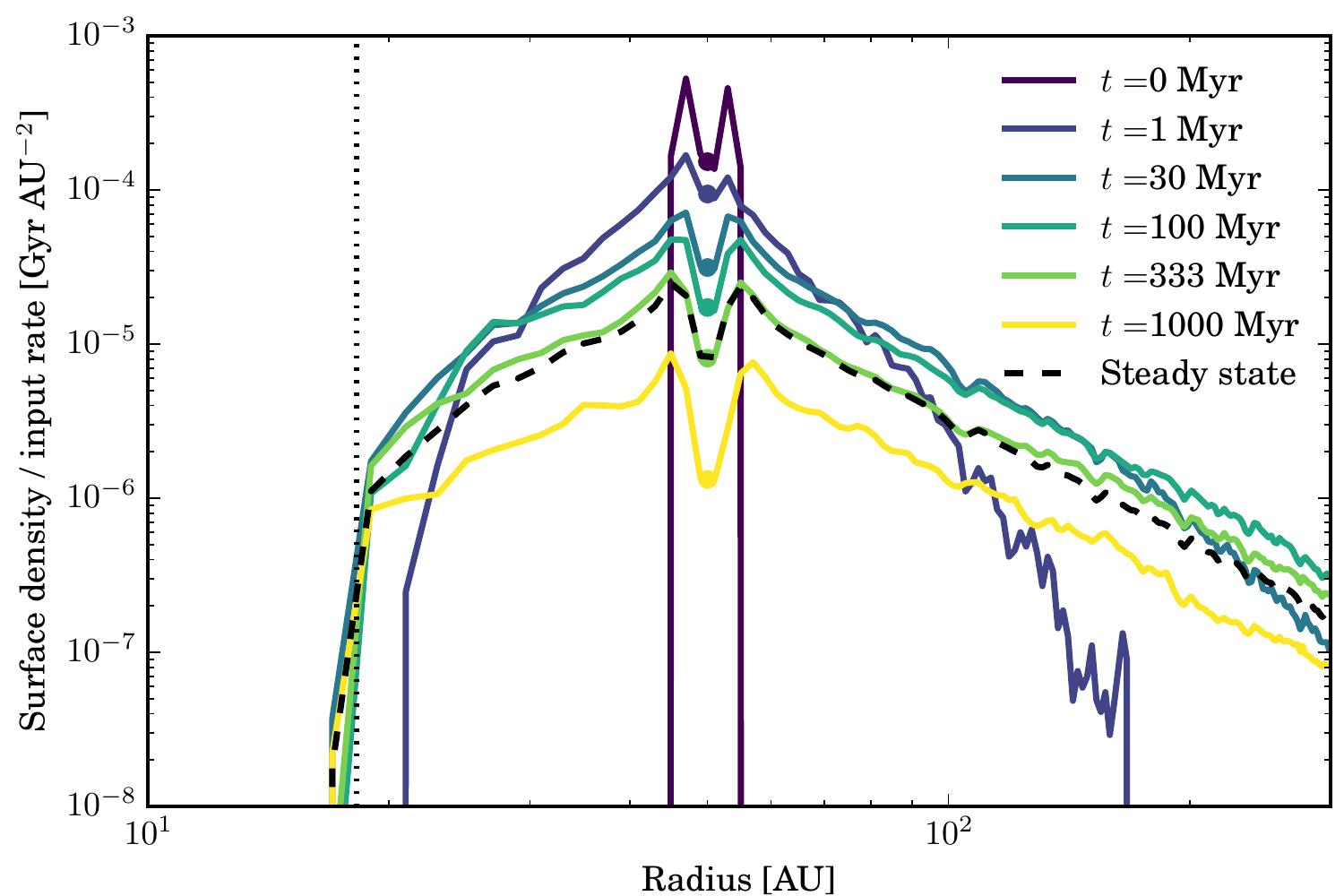}
    \end{subfigure}\\
    \begin{subfigure}{1.0\columnwidth}
      \includegraphics[trim=0.0cm 0.0cm 0.0cm 0.0cm, clip=true, width=1.0\columnwidth]{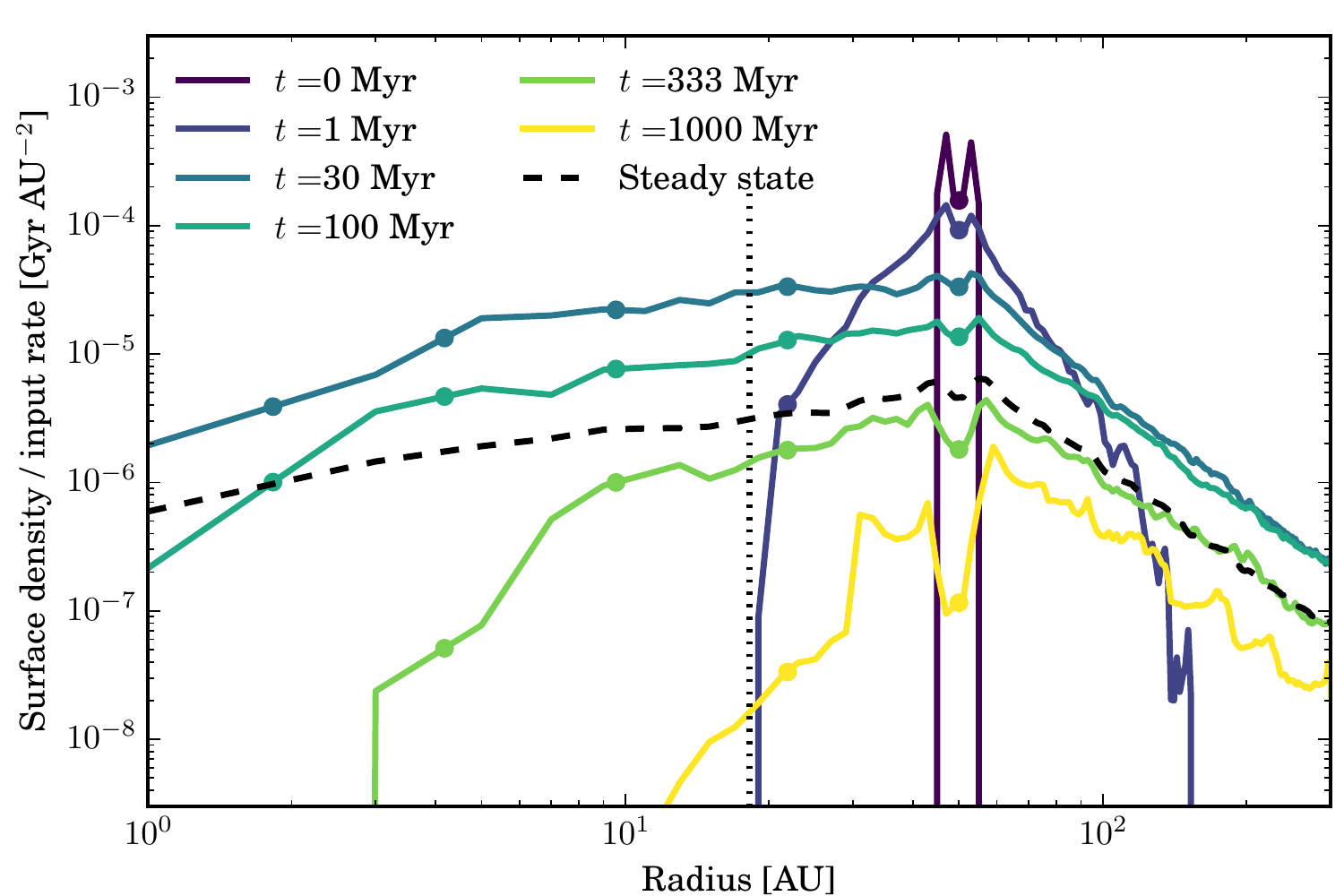}
    \end{subfigure}
  \end{tabular}
 \caption{Surface density of particles as a function of time (colours)
   for the scenario in which particles start at $t=0$ in the chaotic
   zone of a planet at 50 AU in the single planet case (top) and in
   our reference multi-planet system (bottom). The black dashed line
   represents the steady state surface density when particles are
   input in the system at a constant rate in the chaotic zone of the
   outermost planet. The radial location of the dots corresponds to
   the semi-major axis of each planet. The black dotted vertical line
   represents $q_\mathrm{min,p1}$. At $t=0$ the distribution of
   particles is double peaked near 50 AU because we removed those
   particles that stayed as Trojans. Note that the x-axis in the lower
   panel extends further in than in the upper panel.}
 \label{fig:srevol}
\end{figure}

\subsection{Single planet system}
\label{sec:res_single}


In the single planet system we find that of the total number of
particles lost, 85\% of particles are ejected, 15\% are accreted by
the planet and none cross the 0.5~AU inner edge. The ejection
timescale is 380 Myr, consistent with the 390 Myr scattering timescale
estimated with Equation \ref{eq:teject}. After 1 Gyr of evolution,
17\% of particles remain in the system on highly eccentric orbits,
most of which have semi-major axes beyond 50 AU. As noted in
\S\ref{sec:analysis}, these fractions exclude Trojans which represent
31\% of the original $10^3$ simulated particles. The top panel in
Figure \ref{fig:srevol} shows the evolution of the surface density of
particles when they all start near the outermost planet at $t=0$
(colours) and the steady state surface density when particles are
input at a constant rate (black). In all cases the surface density
peaks near 50 AU (where particles are initially placed). Beyond 50 AU,
$\Sigma(r)$ decreases steeply with orbital radius $\propto
r^{-\gamma}$, where $\gamma$ is about -3 as expected for a scattered
disc population with a common pericentre \citep{Duncan1987}. Within
50~AU $\Sigma(r)$ decreases towards smaller orbital radii, but with a
sharp edge at 18~AU, which is the location of $q_\mathrm{min,p1}$,
expected since no particles should be scattered interior to this.

\subsection{Reference system}
\label{sec:res_reference}

When considering our reference chain of equal mass 30~\Me\ planets
separated by 20~mutual Hill radii, we find that the fraction of
particles that are ejected increases relative to the single planet
case from 85 to 93\%, with a shorter ejection timescale of 120 Myr
(compared with 380 Myr in the single planet case). The shorter
ejection timescale is due to multiple scattering with interior planets
as well as the outermost planet, which makes particles evolve faster
onto unbound orbits. After 1 Gyr of evolution, only 5\% remain in the
system, most of which are in a scattered disc beyond 50 AU (see see
yellow line on the bottom panel of Figure \ref{fig:srevol}). As
predicted, the presence of multiple planets makes it easier for
particles to be scattered inwards, and the fraction of particles that
cross 0.5~AU increases from $<0.65\%$ to $3.7^{+0.9}_{-0.7}\%$. On the
other hand, the fraction of particles that are accreted by planets
decreases relative to the single planet case from 15\% to 3.6\% (0.4\%
by inner planets). This is because particles are scattered by multiple
planets, thus increasing the level of stirring and reducing the number
of close encounters that they have with the outermost planet (e.g. the
steady state surface density near the outermost planet is lower by a
factor of 2 in the reference system).

\begin{figure*}  
    \centering
    \begin{tabular}[t]{|c|c|}
      \begin{subfigure}{0.5\textwidth}
        \centering
        \includegraphics[trim=0.0cm 0.0cm 0.1cm 0.0cm, clip=true, width=1.0\linewidth]{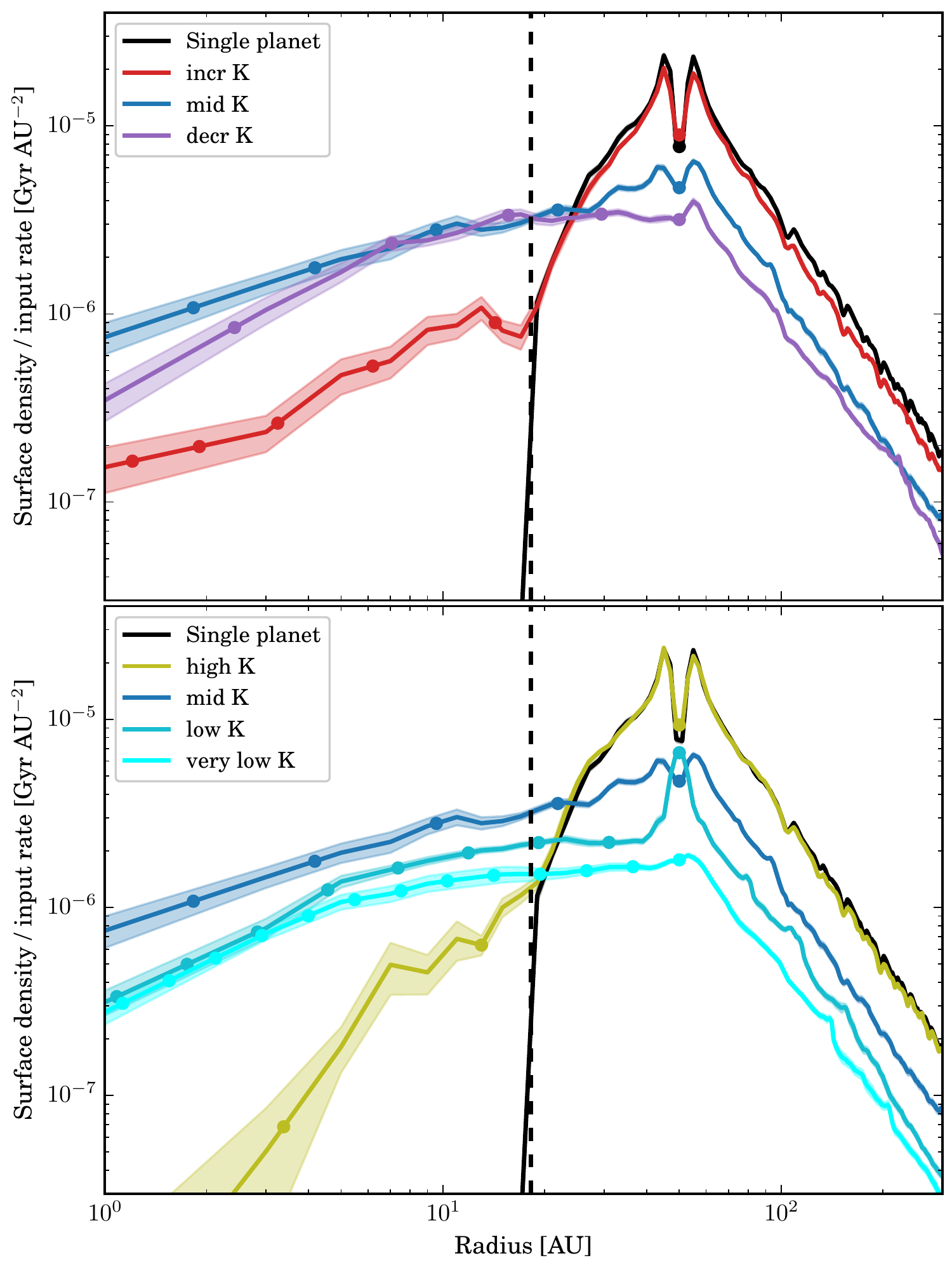}
      \end{subfigure}
      &
      \begin{tabular}{c}
        \begin{subfigure}[t]{0.45\textwidth}
          \centering
          \includegraphics[trim=0.08cm 0.15cm 0.0cm 0.0cm, clip=true,width=1.0\textwidth]{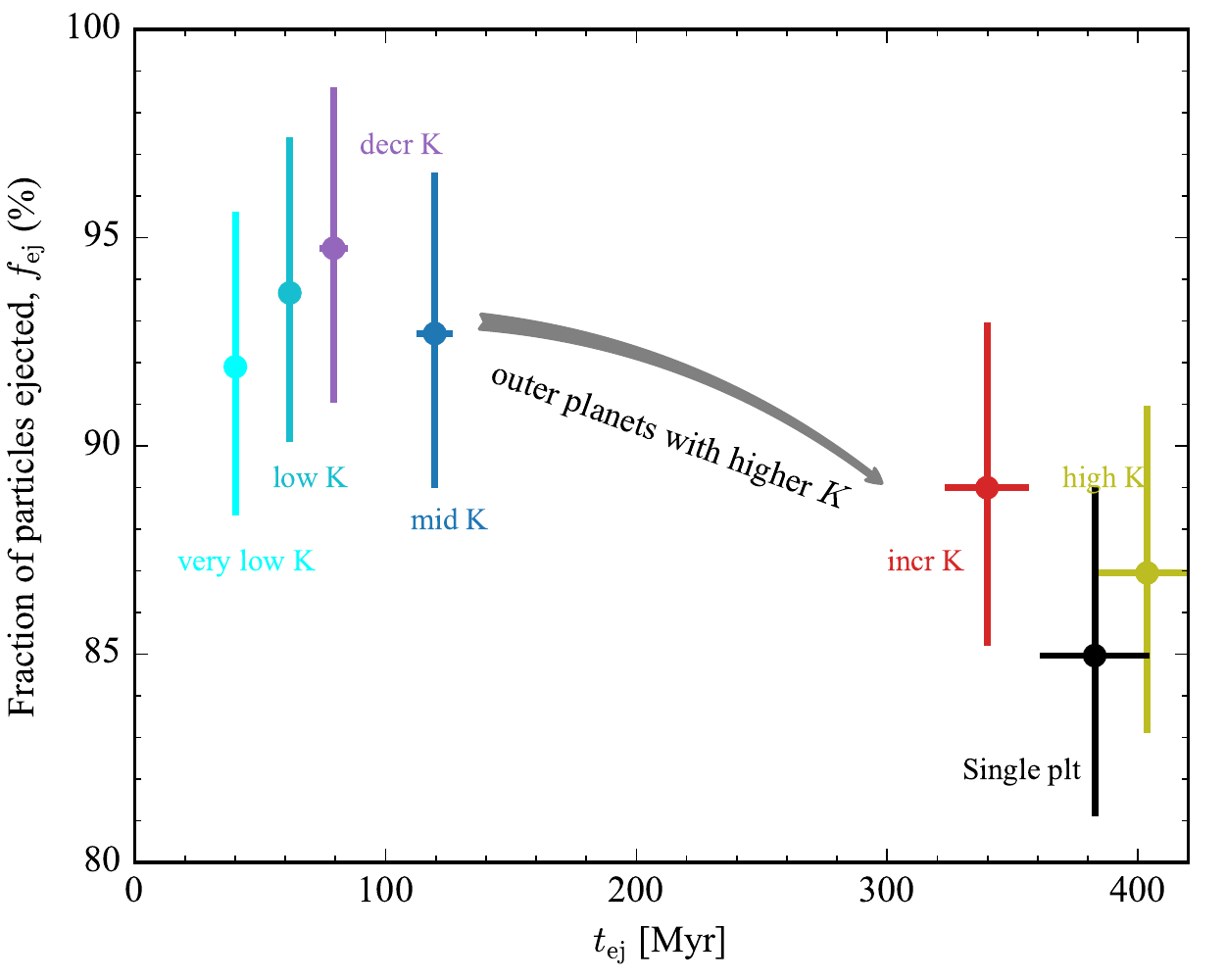}
        \end{subfigure}\\
        \begin{subfigure}[t]{0.45\textwidth}
          \centering
          \includegraphics[trim=0.3cm 0.0cm 0.0cm 0.35cm, clip=true,width=1.0\textwidth]{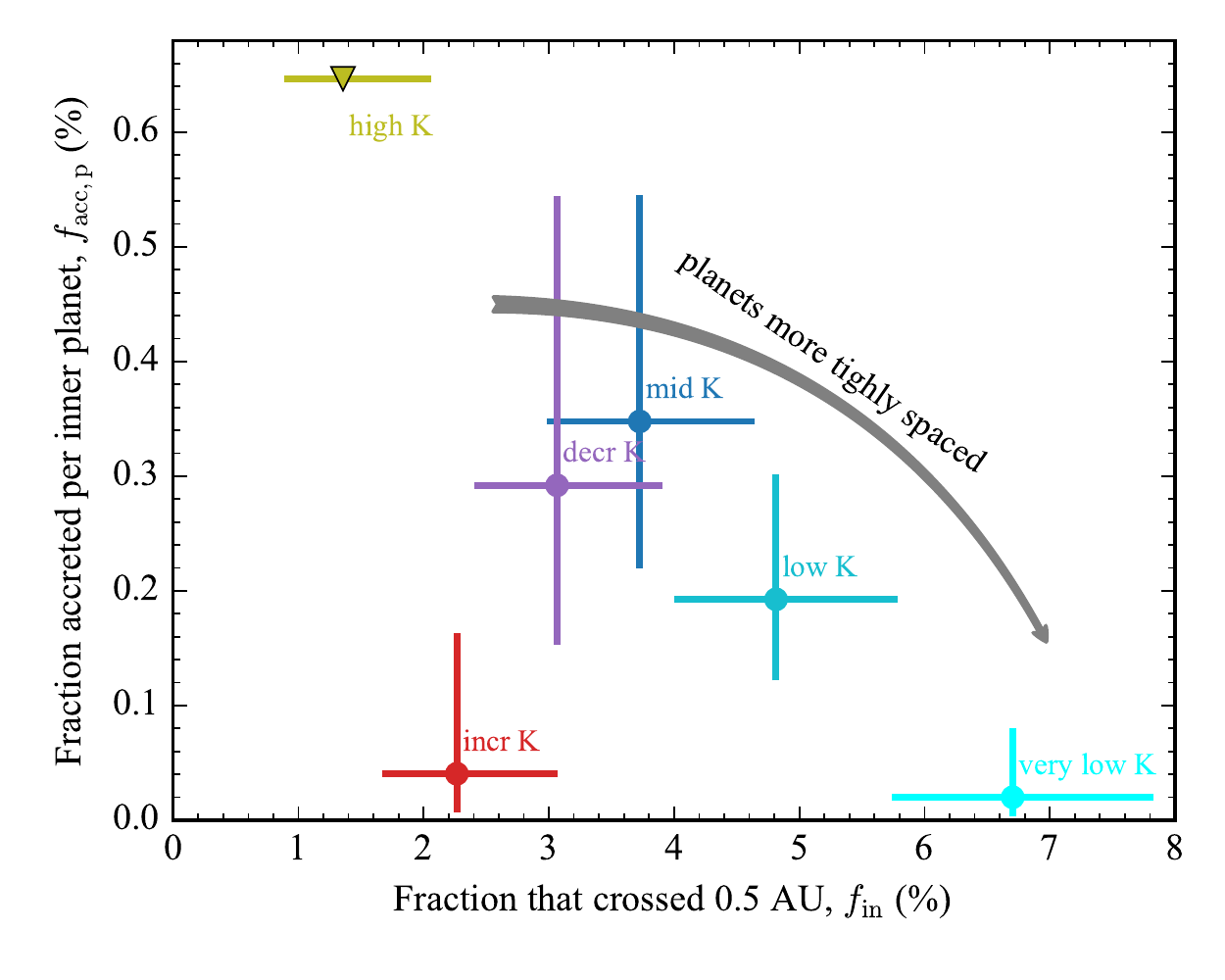}
        \end{subfigure}
      \end{tabular}\\
    \end{tabular}
    \caption{Results from N-body simulations for planet configurations
      with 30~\Me \ planets and varying the spacing between
      planets. \textbf{\textit{Left:}} steady state surface density
      distribution of particles. Models varying $\alphad$ are shown at
      the top left panel, while models with constant separations in
      mutual Hill radii of 8, 12, 20 and 30 at the bottom left.  The
      radial location of the dots corresponds to the semi-major axis
      of each planet. Particles are input in the chaotic zone of the
      outermost planet at a constant rate. The black dashed vertical
      line represents $q_\mathrm{min,p1}$. The black continuous line
      represents the surface density of a single planet system
      ($\aplt=50$~AU). \textbf{\textit{Top right:}} fraction of
      particles ejected vs ejection timescales, calculated as the
      median of epochs at which particles are ejected.
      \textbf{\textit{Lower right:}} fraction of particles accreted
      per planet within 10 AU vs fraction that crosses the inner
      boundary at 0.5 AU.}\label{fig:results_spacing}
\end{figure*}

The bottom panel of Figure \ref{fig:srevol} shows the evolution of
$\Sigma(r,t)$ and its steady state form when particles are input at a
constant rate. Beyond 50 AU, the system has a surface density similar
to the single planet case, but within 50 AU it is flatter and extends
within $q_\mathrm{min,p1}$ as particles are scattered by inner
planets.  Within 10 AU, $\Sigma(r)$ approximates to a power law with a
slope of $\sim0.7$, which flattens out towards 50 AU. This slope is
overall steeper compared to the surface density expected in a P-R drag
scenario \citep{vanLieshout2014, Kennedy2015prdrag}, thus if material
was detected between an outer belt and exozodi, the slope could be
used to discern between scattering and P-R drag scenarios. Note that
the surface density within a few AU could be underestimated due to our
boundary condition at 0.5 AU. Particles within a few AU are likely to
be highly eccentric after being scattered multiple times, with
pericentres that could reach 0.5 AU, and thus, be removed from the
simulation. This is investigated further in
\S\ref{sec:simparameters} decreasing our inner boundary to 0.1
AU. The short lifetime of particles in the system with multiple
planets is also manifested in the surface density at $t=1$~Gyr that is
significantly lower compared to the single planet system. It is also
worth noting that whereas Figure \ref{fig:srevol} shows the evolution
of the surface density, future figures will only show the steady state
surface density.



\subsection{Varying $K(a)$}
\label{sec:res_spacing}

Here we present results from four simulations of equal mass planets
(30~\Me), but with different spacing, measured with the number of
mutual Hill radii ($K$) between adjacent planets. These are planet
configurations 4, 5, 8, 9 and 11, for which $\alphad$ and $K_0$ vary
(see Figure \ref{fig:setups} upper and lower right panels). We also
include results from the reference and single planet system for
comparison. Figure \ref{fig:results_spacing} presents the results for
$\Sigma(r)$, $\fej$, $\tej$, $\fin$ and $\faccplt$.


\subsubsection{Varying $K$ uniformly}

When increasing the spacing from $K=20$ (reference system, dark
blue, called mid K in Figure \ref{fig:results_spacing}) to 30 (yellow, high K)
we observe the following effects:
\begin{enumerate}
 \item The surface density of scattered particles is higher in the
   outer scattered disc, but much lower towards smaller orbital
   radii. In fact, the surface density is very similar to the single
   planet case beyond $q_\mathrm{min, p1}$. Only a very small fraction
   extends within 18 AU as expected because the next planet in the
   chain is inside $q_\mathrm{min, p1}$.
\item The fraction of ejected particles is 87\%, slightly lower than
  the 93\% in our reference system and similar to the single planet
  case. This difference is due to the higher fraction of accreted
  particles (12\% by the outermost planet).
\item The ejection timescale is also increased from 120 to 404
  Myr. The longer lifetime of particles causes the surface density to
  be higher beyond 50~AU in the case of large planet spacing. In fact,
  both $\fej$ and $\tej$ are very similar compared to the single
  planet case. The rest of the particles that are not ejected are
  mostly accreted by the outermost planet (12\%).
\item Only a very small fraction of particles are able to cross 0.5 AU
  (1.4\%).
\item The fraction accreted per inner planet is zero, but
  statistically consistent with the results of our reference system
  ($\lesssim0.65\%$). This is because there is only one planet within
  10~AU in this system, therefore there is a large uncertainty on
  $\faccplt$.
\item After 1 Gyr of evolution 14\% of particles remain in the
  scattered disc beyond the outermost planet.

\end{enumerate}

Therefore, we find that when increasing the spacing, the results
approximate to the single planet case, with the subtle difference that
a few particles are able to be scattered within $q_\mathrm{min, p1}$.
These results confirm the first prediction, that planet configurations
with outer planets too widely spaced, i.e. with
$\apltii<q_\mathrm{min,p1}$, would be inefficient at scattering
particles inwards as multiple scattering is hindered.

When $K$ is lower, e.g. 12 or 8 (represented with light blue colours)
instead of 20, we observe the following effects:
\begin{enumerate}
  \item The surface density of the scattered disc is lower at both
    small and large orbital radii compared to our reference planet
    configuration.
  \item On the other hand, $\fin$ increases as the planet separation
    ($K$) is decreased. This is because scattering by inner planets
    becomes more likely as particles do not require high
    eccentricities to reach the next planet in the chain. Therefore,
    inward scattering happens faster and before particles get a kick
    strong enough to be ejected from the system or get accreted by a
    planet. Note that although the amount of material being passed
    inwards is higher, $\Sigma(r)$ is lower. This is because the
    steady state surface density is also proportional to the lifetime
    of particles at a specific orbital radius.  For example, at 10 AU we find
    that $\Sigma(r)$ is lower compared to the reference system by a
    factor that is consistent with the ratios of ejection timescales
    (see Table \ref{tab:results}), which is a proxy for the lifetime
    of particles in the scattered disc. We also rule out that
      the decrease in $\Sigma(r)$ is due to the increase in $\fin$, as
      the removal of particles only has an effect within 4~AU (see
      appendix \ref{sec:inneredge}).
  \item We also find a decrease in the fraction of particles accreted
    per inner planet when the planet separation becomes low ($K=12$
    and 8, light blue points in lower right panel in Figure
    \ref{fig:results_spacing}). This effect is also observed in
    \S\ref{sec:res_mass} at a significant level. This could be due to
    the lower surface density as the planet masses are the same. Other
    factors could be at play too, such as the eccentricity and
    inclination distribution of particles which can change the
    distribution of relative velocities in close encounters (Equations
    \ref{eq:Racc} and \ref{eq:crosssection}). In fact, we do find that
    in systems with low spacing the distribution of eccentricities and
    inclinations is slightly shifted towards higher values by a few
    percent compared to the reference system, although this difference
    is not large enough to explain the lower accreted
    fraction. Therefore, we conclude that it is the change in surface
    density the main factor at decreasing the accretion onto
    planets. These results suggest that there is an optimum planet
    spacing that maximises the accretion onto inner planets for
    systems of equal mass planets.
\end{enumerate}


\subsubsection{Varying $K$ as a function of $\aplt$ ($\alphad$)}


When the planet spacing decreases towards larger orbital radii
(purple) the surface density remains very similar to our reference
scenario, although slightly lower between 30-150 AU. This simulation
also has a similar or consistent fraction of particles that are
accreted per inner planet and that cross 0.5~AU compared to our
reference case. On the other hand, the ejection timescale in this
simulation (79~Myr) is more similar to the low spacing case ($K=12$),
which has outermost planets at similar separations. This suggests that
the ejection timescale is dominated by the separation of the two
outermost planets rather than the average separation of planets in the
system (for equal mass planets). However, despite the fact that the
outer planets are close to each other, the fraction of particles that
cross 0.5 AU is lower or consistent with the reference system (dark
blue).

\begin{figure*}  
    \centering
    \begin{tabular}[t]{|c|c|}
      \begin{subfigure}{0.5\textwidth}
        \centering
        \includegraphics[trim=0.0cm 0.0cm 0.0cm 0.1cm, clip=true,width=1.0\linewidth]{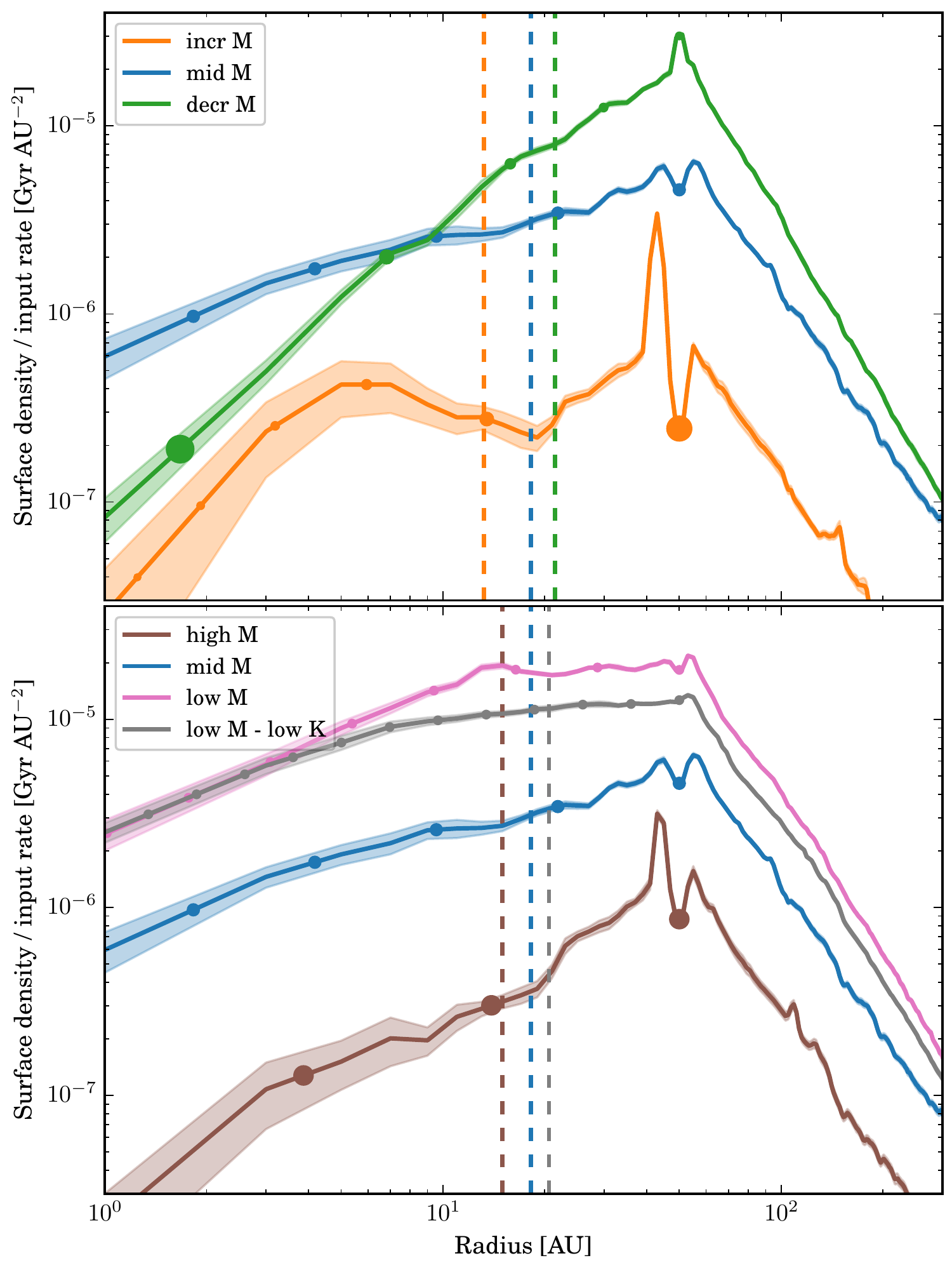}
      \end{subfigure}
      &
      \begin{tabular}{c}
        \begin{subfigure}[t]{0.45\textwidth}
          \centering
          \includegraphics[trim=0.1cm 0.15cm 0.0cm 0.0cm, clip=true,width=1.0\textwidth]{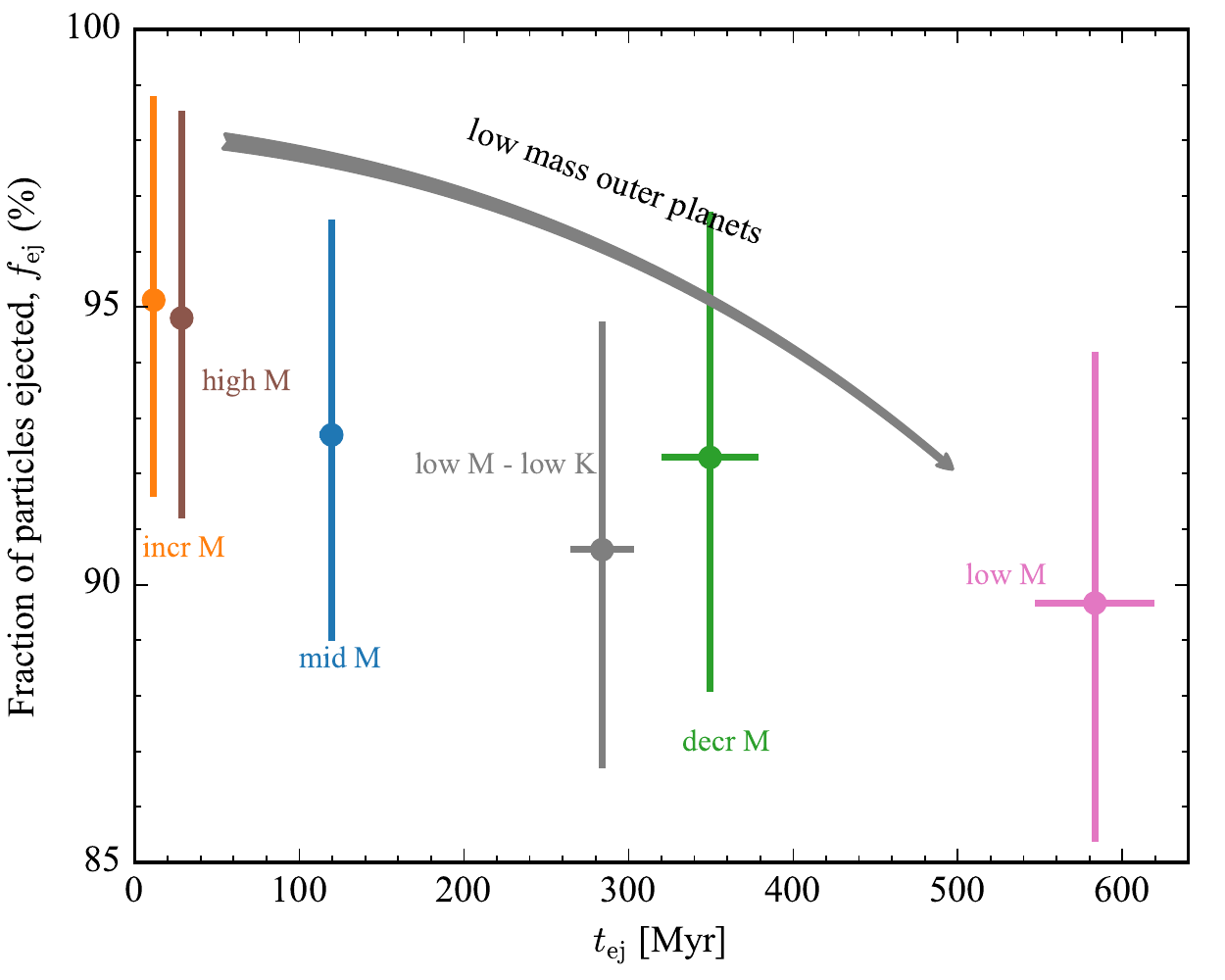}
        \end{subfigure}\\
        \begin{subfigure}[t]{0.45\textwidth}
          \centering
          \includegraphics[trim=0.3cm 0.0cm 0.0cm 0.35cm, clip=true,width=1.0\textwidth]{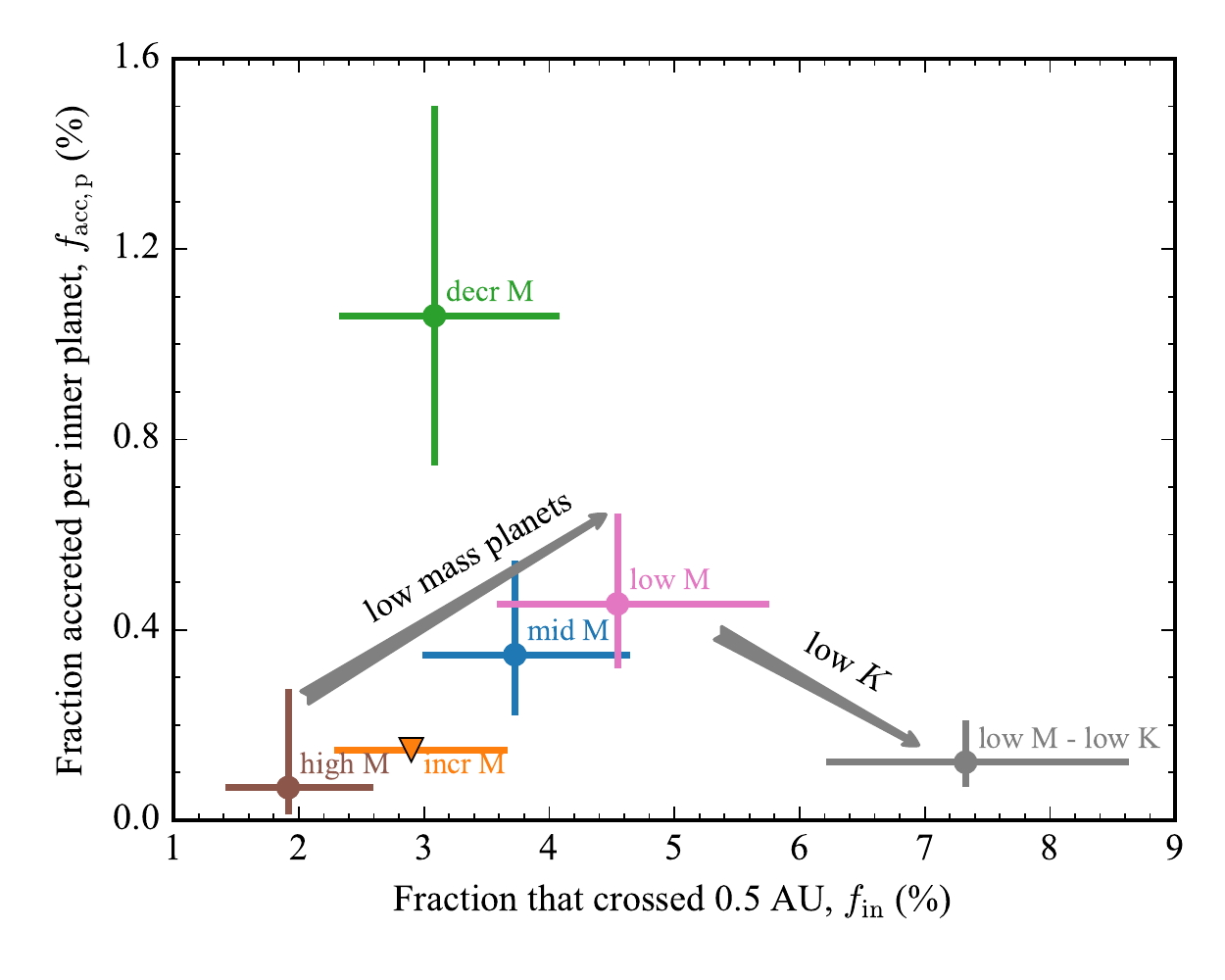}
        \end{subfigure}
      \end{tabular}\\
    \end{tabular}
    \caption{Results from N-body simulations for planet configurations
      with a constant planet spacing of 20 mutual Hill radii ($K=20$)
      and varying their masses. \textbf{\textit{Left:}} steady state
      surface density distribution of particles. Models varying
      $\alpham$ are shown at the top left panel, while models with
      constant masses of 10, 30 and 90 \Me \ at the bottom left.  The
      radial location of the dots corresponds to the semi-major axis
      of each planet and their size representing their masses with an
      arbitrary scale. Particles are input in the chaotic zone of the
      outermost planet at a constant rate. The dashed vertical lines
      represent $q_\mathrm{min,p1}$. \textbf{\textit{Top right:}}
      fraction of particles ejected vs ejection timescales, calculated
      as the median of epochs at which particles are ejected.
      \textbf{\textit{Lower right:}} fraction of particles accreted
      per planet within 10 AU vs fraction that crosses the inner
      boundary at 0.5 AU. The inverted triangle represents a $2\sigma$
      upper limit. The grey line (lower left panel) and points (right
      panels) represent a system with 10 \Me \ planets closely
      separated by 12 mutual Hill radii.}\label{fig:results_mass}
      
\end{figure*}

In the system with increasing planet spacing (red) we find that the
ejection timescale and surface density beyond $q_\mathrm{min,p1}$ are
very similar to the case of high $K$ (yellow) or the single planet
case. As the spacing of the outermost planets increases scattering is
dominated by the outermost planet. However, the surface density within
$q_\mathrm{min,p1}$ is much higher, as is the fraction of particles
that get to 0.5 AU compared to the high $K$ and single planet
systems. The number of accreted particles per inner planet is lower
compared to our reference system.

Neither increasing nor decreasing the spacing as a function of orbital
radius increases the fraction of particles that get into the inner
regions compared to our reference system. However, if we compare
planet configurations with similar spacing for their outer planets
(i.e. pairs yellow-red and light blue-purple) we find that if the
planet spacing decreases towards smaller orbital radii, then the fraction of
particles that cross 0.5 AU and the surface density within
$q_\mathrm{min,p1}$ increases. Therefore, we predict that a system
with both $\apltii>q_\mathrm{min,p1}$ and $\alphad\geq0$ will be very
efficient at passing material inwards.


\subsubsection{Conclusions regarding planet spacing}
Therefore based on these results we conclude the following:
\begin{enumerate}
  \item For efficient inward multiple scattering, outer planets must
    be close enough such that $\apltii>q_\mathrm{min,p1}$---noting
    that $q_\mathrm{\min,p1}$ depends on the initial condition of
    particles which could be different to that assumed here.
  \item Closely spaced outer planets means that particles are ejected
    faster---assuming particles start in the outer regions of the
    system.
  \item Planets in compact configurations, with either uniform spacing
    or decreasing towards smaller orbital radii for a fixed outermost planet
    separation, are most efficient at passing particles inwards.
  \item The more closely packed the planets are, the lower the surface
    density in the scattered disc, and the fewer impacts per inner
    planet.
\end{enumerate}

\subsection{Varying $\Mplt(a)$}
\label{sec:res_mass}

In this section, we present results from four other simulations of
planets separated by 20 mutual Hill radii, but varying their
masses. These are planet configurations 2, 3, 6 and 7, for which
$\alpham$ and $M_0$ vary. We also include the reference system and
configuration 10 which has low mass planets closely spaced to see the
effect of decreasing both planet masses and spacing. Figure
\ref{fig:results_mass} presents the results for $\Sigma(r)$, $\fej$,
$\tej$, $\fin$ and $\faccplt$.


\subsubsection{Varying $\Mplt$ uniformly}
\label{sec:varM}
Our simulations show that at constant separation ($K$), the surface
density, the ejection timescale, and the fractions of particles that
are accreted per inner planet and that cross the inner edge, decrease
when increasing planet mass (Figure \ref{fig:results_mass}, see
changes from pink to dark blue and from dark blue to brown). The
ejection timescale decreases in factors of $\sim4-5$ when increasing
$\Mplt$ from 10 to 30 \Me\ and from 30 to 90 \Me, as higher mass
planets scatter on shorter timescales. Note that this factor is
smaller than that predicted using Equation \ref{eq:teject} which could
be due to the lower number of planets when increasing $\Mplt$ (while
keeping the spacing constant in mutual Hill radii). The change in
ejection timescale also results in the surface density increasing by
similar factors when decreasing $\Mplt$. These strong differences in
surface density cause also an increment in the fraction of particles
accreted per inner planet when decreasing $\Mplt$. We also find that
all inner planets accrete a similar fraction of particles in the
system of low mass planets (pink). While 31\% of particles remain in
the system by 1 Gyr, when extending the integration to 5 Gyr we find
that the fractions of particles ejected, accreted and that cross 0.5
AU do not change significantly (see \S\ref{sec:simparameters}).

We can use these results to derive an empirical relation for the
fraction of particles that get into 0.5 AU and surface density as a
function of the planet mass, for a fixed input rate and a chain of
equal mass planets separated by 20 mutual Hill radii. We find that
$\fin$ varies approximately as $\Mplt^{-0.37\pm0.13}$ and the surface
density at 10 AU as $\Mplt^{-1.6\pm0.2}$. This indicates that the
surface density in the scattered disc is not directly proportional to
the fraction of particles that get to 0.5 AU, with the surface density
being more sensitive to changes in planet mass. We can combine these
two empirical expressions to find $\fin\propto\stenau^{0.23\pm0.09}$
for chains of equal mass planets.

\subsubsection{Varying $\Mplt$  as a function of $\aplt$ ($\alpham$)}

When varying $\alpham$ with a fixed planet mass of 30~\Me \ at 10~AU
(see top left panel of Figure \ref{fig:setups} and compare orange blue
and green) we find the following. For planet masses decreasing with
orbital radius ($\alpham=-1$, green) the ejection timescale is
increased relative to the reference system. Similarly, for planet
masses increasing with orbital radius ($\alpham=1$, orange) the
ejection timescale is decreased. The changes in the ejection timescale
are due to its strong dependence on the mass of the outermost planet
as that is where particles are initiated. This is similar to what we
found in \S\ref{sec:res_spacing} in which the ejection timescale is
dominated by the separation of the outermost planets.

We also find that the surface density changes significantly when
varying $\alpham$ (Figure \ref{fig:results_mass} top left panel). For
planet masses decreasing with distance to the star (decr M, green),
the surface density becomes steeper both inside and outside 50 AU,
with a slope of $\sim1.5$ within 50 AU. This causes $\Sigma(r)$ to be
lower (higher) within (beyond) 10 AU compared to our reference
system. For planet masses increasing with distance to the star, the
surface density is lower and flatter (incr~M, orange). The steep slope
within 4 AU and the peak at 43 AU seen in the orange line are due to
our inner boundary (together with low number statistics) and particles
trapped in the 5:4 mean motion resonance with the outermost planet,
respectively. The slope of the surface density within 50~AU depends
strongly on $\alpham$ because the lifetime of particles or ejection
timescale is a function of $\Mplt$ and $\aplt$ (Equation
\ref{eq:teject}), hence, more negative $\alpham$'s will result in
steeper positive slopes.

Regarding the efficiency of particles reaching 0.5 AU ($\fin$), the
bottom right panel in Figure \ref{fig:results_mass} shows that this is
slightly lower compared to the reference system for both $\alpham=-1$
and 1 (decr~M and incr~M, respectively), but consistent within
errors. However, when comparing systems with outermost planets of
similar mass (green-pink and brown-orange) we find that positive
$\alpham$ results in a slightly higher fraction of particles reaching
0.5 AU, while slightly lower for negative $\alpham$. This partially
contradicts one of our predictions in \S\ref{sec:analytic}, that
chains of planets with mass decreasing with distance to the star
(i.e. scattering timescales increasing with distance) would be better
at inward scattering than systems with mass increasing with
distance. We expect that this is due to the system with planet masses
decreasing with distance from the star (green) having inner planets in
the ejection regime (see discussion in
\S\ref{sec:predictionscorrect}).

\begin{figure}
  \centering \includegraphics[trim=0.0cm 0.0cm 0.0cm 0.0cm, clip=true,
    width=1.0\columnwidth]{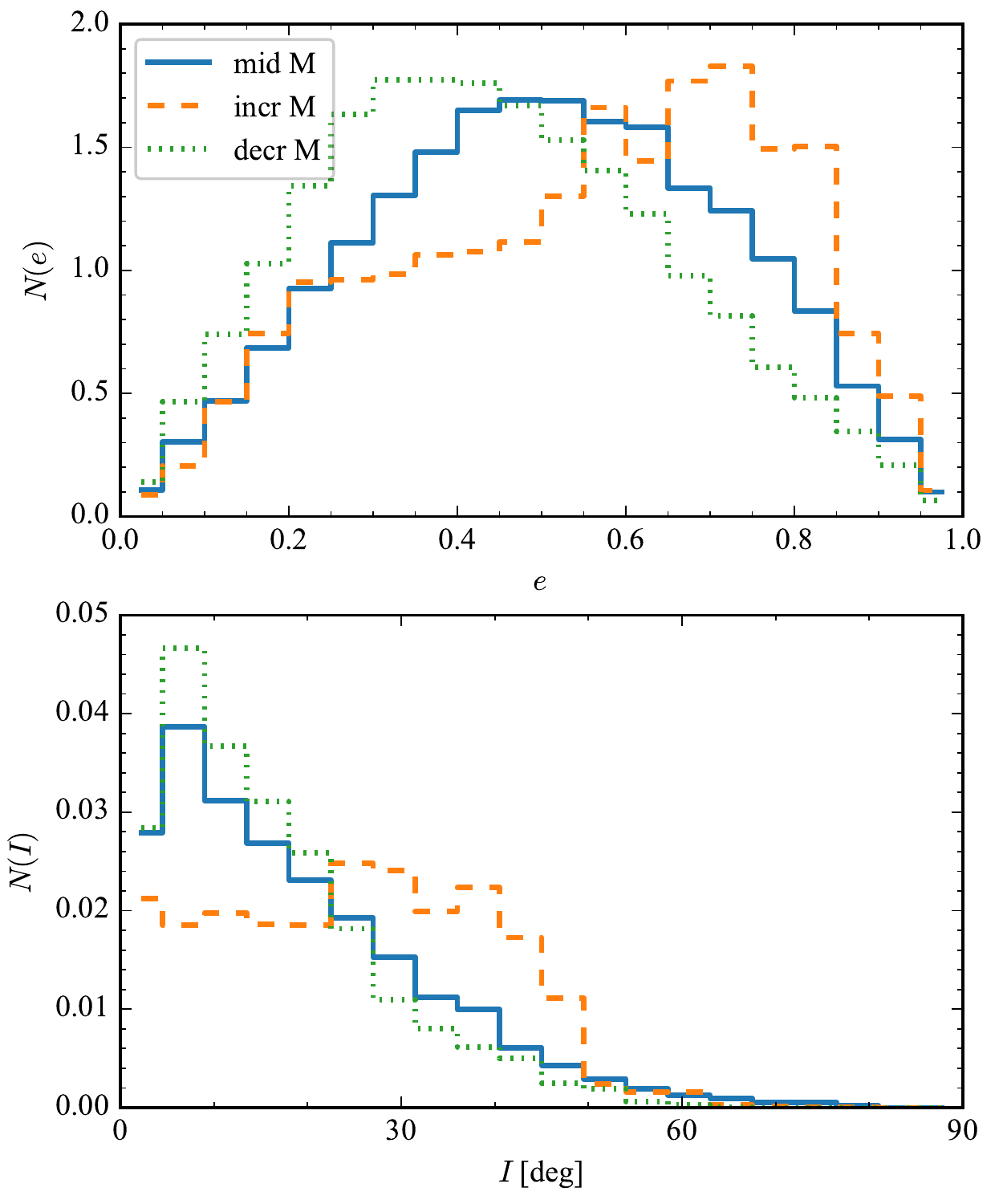}
 \caption{Normalized distribution of eccentricities and inclinations
   of particles having close encounters with planets within 10 AU.}
 \label{fig:eidist}
\end{figure}

We find that the fraction of particles accreted per inner planets is
the highest in the system with planet masses decreasing with orbital
radius (0.8 and 1.3\% for the planets at 2 and 7 AU, respectively)
compared to the other 11 simulated systems. This is surprising as it
has one of the lowest surface densities within 10 AU. On the contrary,
no particles were accreted by inner planets in the system with planet
masses increasing with orbital radius ($\faccplt\lesssim0.15\%$). To
understand what drives the higher accretion fraction per inner planet
in the configuration with decreasing mass with orbital radius (green),
we need to consider the density of particles around their orbits, the
planet masses and radii, and the relative velocities of particles that
the inner planets encounter (Equation \ref{eq:Racc}). As seen in
Figure \ref{fig:results_mass} the surface density is not particularly
high within 10 AU compared, for example, to our reference
configuration. Moreover, we analyse the number of close encounters per
planet (those that get within 3 mutual Hill radii) finding that the
surface density is a reasonable tracer of this quantity, being lower
within 10 AU. The higher level of accretion of the innermost planet
for the simulation with decreasing mass with orbital radius (green)
compared to the innermost planet in our reference simulation (dark
blue) could be explained mostly by its higher planet mass (180 vs
30~\Me) and radius. However, the second innermost planet also accretes
significantly more than the inner planets in our reference system
($\gtrsim3$ times more), but with a slightly lower $\Sigma(r)$ and a
similar planet mass of 44~\Me and radius (remember that $\Gamma$ is
proportional to $\Mplt^{4/3}$ for these planet masses and semi-major
axes). Therefore, the fraction of accreted particles must be also
enhanced by a difference in the distribution of eccentricities and
inclinations of particles in these regions, which defines the relative
velocities at which these planets are encountering scattered particles
and the collisional cross sections (see Equation
\ref{eq:crosssection}). When comparing the distribution of
eccentricities and inclinations of particles having close encounters,
we find that their distributions are indeed significantly shifted
towards lower values for the decr M system (see Figure
\ref{fig:eidist}). Therefore encounters happen on average at lower
relative velocities and in a flatter scattered disc (lower $h$). The
lower eccentricities and inclinations, and higher collisional
cross-section can increase $R_\mathrm{acc,plt}$ by a factor of $\sim3$
for the second innermost planet of the decreasing mass system (green)
compared to inner planets in our reference system, which is consistent
with the higher number of impacts that we find for close-in
planets. The lower eccentricities and inclinations are likely due to
the fact that the outer planets scattering the material in from 50~AU
are lower in mass. We observe a similar effect for the system with
equal low mass planets (pink) and the opposite for the incr M system,
i.e higher eccentricities and inclinations are likely due to the high
mass of the outermost planets.

\subsubsection{Low planet mass and spacing}
Finally, in order to test if lowering the spacing of low mass planets
can increase even more $\fin$ and possibly reduce $\faccplt$ as found
in \S\ref{sec:res_spacing}, in Figure \ref{fig:results_mass} we
overlay a system with 10~\Me\ planets spaced with $K=12$ (grey). In
accordance with our findings in \S\ref{sec:res_spacing}, we find that
such a system has a higher number of particles that get to 0.5 AU
(7\%), but significantly lower $\faccplt$. This is due to two effects:
the surface density of the scattered disc is lower, but also we find
that the distribution of eccentricities and inclinations in close
encounters is shifted towards higher values. Thus, it could be that
there is an optimum separation ($>8\Rhillm$) that maximises the amount
of accretion per inner planet.

\subsubsection{Conclusions regarding planet masses}

Therefore based on the results varying the masses of the planets we
conclude that:
\begin{enumerate}
\item The surface density decreases with increasing $\Mplt$ for fixed
  planet spacing $K$.
\item The slope of the surface density profile varies when varying the
  planet masses as a function of semi-major axis. For planet masses
  decreasing as a function of orbital radius, $\Sigma(r)$ increases steeply
  with orbital radius compared to our reference system.
\item The fraction of particles that get to 0.5 AU increases when
  decreasing $\Mplt$. Placing the planets closer together increases
  even more the fraction of particles that get to 0.5~AU, but reduces
  the surface density and thus the fraction of particles accreted per
  inner planet.
\item Systems with innermost planets in the ejection regime are less efficient at
  transporting material within 0.5 AU (green and brown).
\item The number of particles accreted per inner planet increases with
  decreasing $\Mplt$ as the surface density is increased due the
  longer lifetime of particles in the disc. However, systems with
  $\alpham<0$ have the highest $\faccplt$ because of the high
  collisional cross-section of inner planets caused by their high mass
  and planet radii, and the low relative velocities of particles in
  the scattered disc.
\end{enumerate}

\subsection{Comparison with our predictions}
\label{sec:predictionscorrect}

In this section we compare our results with previous work from which
we made some predictions in \S\ref{sec:predictions}. From analytic
arguments three main results were expected. First, if particles start
from a cold disc near the outermost planet, then systems with widely
spaced planets, or at least with outermost planets too far from each
other, would have trouble scattering particles inwards. In our
simulations we find that this is approximately true, since systems
with wide outer planets were the ones with the least amount of
material transported to 0.5 AU, confirming results by
\cite{Bonsor2012nbody}.


The second expected outcome was that chains of high mass planets will
scatter and eject particles on shorter timescales. In our simulations
we find that this is true with ejection timescales varying from $580$
to 30~Myr when varying $\Mplt$ from 10 to 90 \Me. Shorter ejection
timescales lead to lower surface densities of scattered particles.

The third expected outcome regarded the mass distribution of the
planets in the chain, with the best systems passing material inwards
being the ones with equal mass chain of planets or decreasing with
distance from the star \citep{Wyatt2017}. We find that this is true
for systems of equal mass planets, being the best of the simulated
systems at scattering material inwards to within 0.5 AU. Systems with
decreasing mass were not as efficient at passing material inward as
they encounter the following problem. The outermost planet must have a
mass high enough that the scattering timescale is shorter than 1 Gyr
(length of simulation), otherwise particles will take too long to be
scattered inwards. This means that if the outermost planet is at 50
AU, then it must be more massive than $\sim10$~\Me\ for a Gyr old
system. However, if the planet masses increase towards smaller orbital
radii, then the innermost planet at 1 AU will be in the ejected region
(see Figure \ref{fig:setups}), hence ejecting most of the particles
instead of scattering them in so that they can reach 0.5 AU. This is
similar to the Solar System where Jupiter ejects most of the minor
bodies that are scattered in from the Kuiper belt, which thus never
get into the inner Solar System.


The fourth expected outcome was that the accretion onto inner planets
(which has implications for the delivery of volatile-rich material
formed in an outer belt) will be higher if they are more massive, the
surface density of particles around their orbits is higher, and
particles are on low eccentricity and inclination orbits. We confirmed
this, however, as expected we also found that these factors are not
independent of each other as the planet mass and spacing affects both
the surface density and the distributions of eccentricity and
inclination of scattered particles. We found that systems of low mass
planets have the highest accretion per inner planet as the density of
particles is highest for these systems. If planet masses are allowed
to vary as a function of orbital radius, we find that the system with low mass
outer planets and high mass inner planets has the highest accretion
per inner planet, as it is an optimum of the different factors
presented above confirming our predictions. Finally, we also found
that there must be an optimum planet separation for delivering
material to inner planets. While particles in systems with planets
widely spaced are less likely to be scattered inwards, particles in
systems of tightly packed planets are ejected on shorter timescales
and have higher relative velocities, hence both wide and small planet
spacing hinder accretion onto inner planets.

\section{Discussion}
\label{sec:discussion}

In this paper we have explored the process of inward scattering for a
variety of planetary systems. Below we discuss our main findings
regarding the surface density of scattered particles, inward transport
to 0.5 AU, and delivering material to inner planets. We also discuss
some of our model assumptions and simulation parameters.

\subsection{Can we detect the scattered discs in a system with an exozodi?}
\label{sec:detectability}


Consider a system that is observed to have an exozodi. Infrared
observations of the dust emission can be used to infer the rate at
which mass is lost from the exozodi, $R_\mathrm{zodi}$. For example,
for $\sim$1~Gyr old systems with exozodis such as Vega, $\eta$~Corvi
and HD69830 this is
$R_\mathrm{zodi}\sim10^{-11}-10^{-9}$~\Me~yr$^{-1}$ \citep[see
  Equation 29 in][]{Wyatt2007hotdust}, being highest for Vega and
$\eta$~Corvi. Here we consider what constraints the results from
\S\ref{sec:results} place on the possibility that an exozodi is
replenished by scattering of planetesimals from an outer belt. The two
conditions that we must consider are whether mass is passed in at a
sufficient rate to replenish the exozodi at a rate $R_\mathrm{zodi}$
and whether this requires the presence of a scattered disc between the
outer disc and the exozodi that is bright enough to be
detectable. Below we first present some general considerations to
estimate the surface density of dust in the disc of scattered
particles to compare it with sensitivity limits from different
telescopes (\S\ref{sec:generalconsiderations}). Then we apply these to
a specific system (\S\ref{sec:etacorvi}).

\subsubsection{General considerations}
\label{sec:generalconsiderations}

As discussed in \S\ref{sec:framework}, here we assume that solids from
an outer belt of planetesimals are input near an outermost planet
(located at 50 AU) at a constant rate $R_\mathrm{in}$, and that when
they cross 0.5 AU their mass is incorporated in a collisional cascade
where solids are ground down to dust. Then $R_\mathrm{zodi}=\fin
R_\mathrm{in}$, and if we assume that scattering has been ongoing for
the system's whole lifetime, the total mass that has been scattered
from the outer belt is $t_{\star}R_\mathrm{zodi}/\fin$, where
$t_{\star}$ is the age of the system. It is worth noting that although
$\fin$ varies as a function of the planetary system architecture, it
does not change by more than an order of magnitude when varying
considerably the planet masses and spacing (based on our simulations,
$\fin$ is in the range 1-7\%). This means that for $\sim1$ Gyr~old
systems with a high $R_\mathrm{zodi}$ of $\sim10^{-9}$~\Me~yr$^{-1}$
(e.g. Vega or $\eta$~Corvi) the amount of material scattered from the
outer belt over the lifetime of the system (assuming this is the
origin of the exozodi) is probably as high as $\sim$10~\Me\ if this
has been a continuous process or even higher if not all of the mass
that reaches 0.5~AU ends up in the exozodi. Note that this is a
significant fraction of estimated masses of their outer belts assuming
a size distribution with a power index of -3.5 and a maximum
planetesimal size of 10-100~km \citep{Marino2017etacorvi,
  Holland2017}.

\begin{figure}
  \centering \includegraphics[trim=0.0cm 0.0cm 0.0cm 0.0cm, clip=true,
    width=1.0\columnwidth]{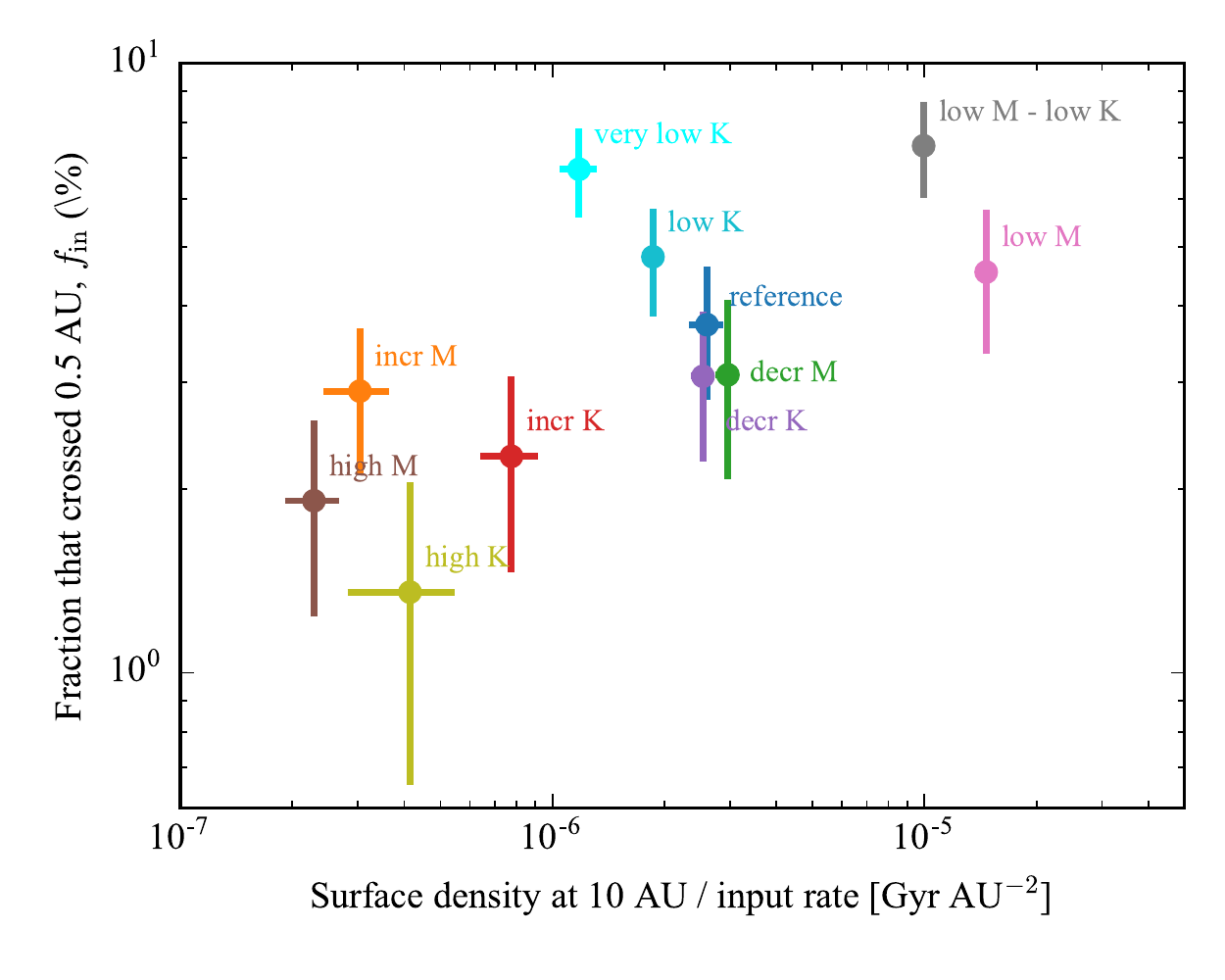}
 \caption{Fraction of particles that cross 0.5 AU vs mass surface
   density at 10 AU divided by the mass input rate for the simulated
   systems presented in Table \ref{tab:setups}.}
 \label{fig:finvsSr}
\end{figure}

Comparing the fraction of particles that get to 0.5 AU with the
surface density of particles (Figure \ref{fig:finvsSr}), we find that
there is a correlation between the two. This correlation is similar to
the one that we found in \S\ref{sec:varM}, with
$\fin\propto\stenau^{0.23\pm0.09}$. That is, while the surface density
varies by two orders of magnitude, $\fin$ only varies by a factor of a
few.  Moreover, although there is a correlation for systems with equal
planet spacing ($K=20$) and different planet masses (high M, reference
and low M), there is an important dispersion for the rest of the
architectures explored. This means that bright exozodis (i.e. high
$\fin$) do not necessarily require a bright scattered disc between the
exozodi and outer belts (i.e. large $\stenau$) if, for example,
massive planets or tightly packed medium mass planets are
present. Therefore, for a given exozodi, upper limits on the amount of
material between these regions can help us to constrain the mass and
separations of intervening planets (see \S\ref{sec:etacorvi}).

To estimate if the scattered disc between the outer belt and the
exozodi could be detected, we assume that the distribution of mass
shown in figures \ref{fig:results_spacing} and \ref{fig:results_mass}
is also representative of the distribution of dust (i.e neglecting
radiation forces acting on small grains). Although collisions can
deplete dust densities through collisions before being lost from the
system (ejected, accreted or transported within 0.5~AU), dust should
also be replenished by collisions of bigger bodies that are also
scattered and have lifetimes longer than the scattering timescales. We
first obtain the total surface density of solids by scaling the
surface density to the necessary input rate ($R_\mathrm{in}$) that can
sustain a given exozodi ($R_\mathrm{zodi}=\fin R_\mathrm{in}$). Then,
we scale the surface density of solids to consider only the mass in
dust grains smaller than 1 cm (as infrared observations at wavelengths
shorter than 1~mm are only sensitive to emission from dust grains
smaller than $\sim1$~cm). In the scenario that we are considering
solids originate in an outer debris belt, thus we assume a standard
-3.5 power law size distribution of solids with a maximum size of
100~km and a minimum size of 1~$\mu$m, roughly the blow-out size for a
Sun-like star. Then the scaling factor to transform the total mass
into the mass of dust grains smaller than 1 cm is
$\sim3\times10^{-4}$. This factor is approximately the same when using
the resulting size distribution of solids at 50 AU after 1 Gyr of
evolution, and taking into account a size dependent disruption
threshold of solids \citep[see middle panel of Figure 9 in
][]{Marino201761vir}.

\begin{figure}
  \centering
 \includegraphics[trim=0.4cm 0.5cm 0.4cm 0.3cm, clip=true, width=1.0\columnwidth]{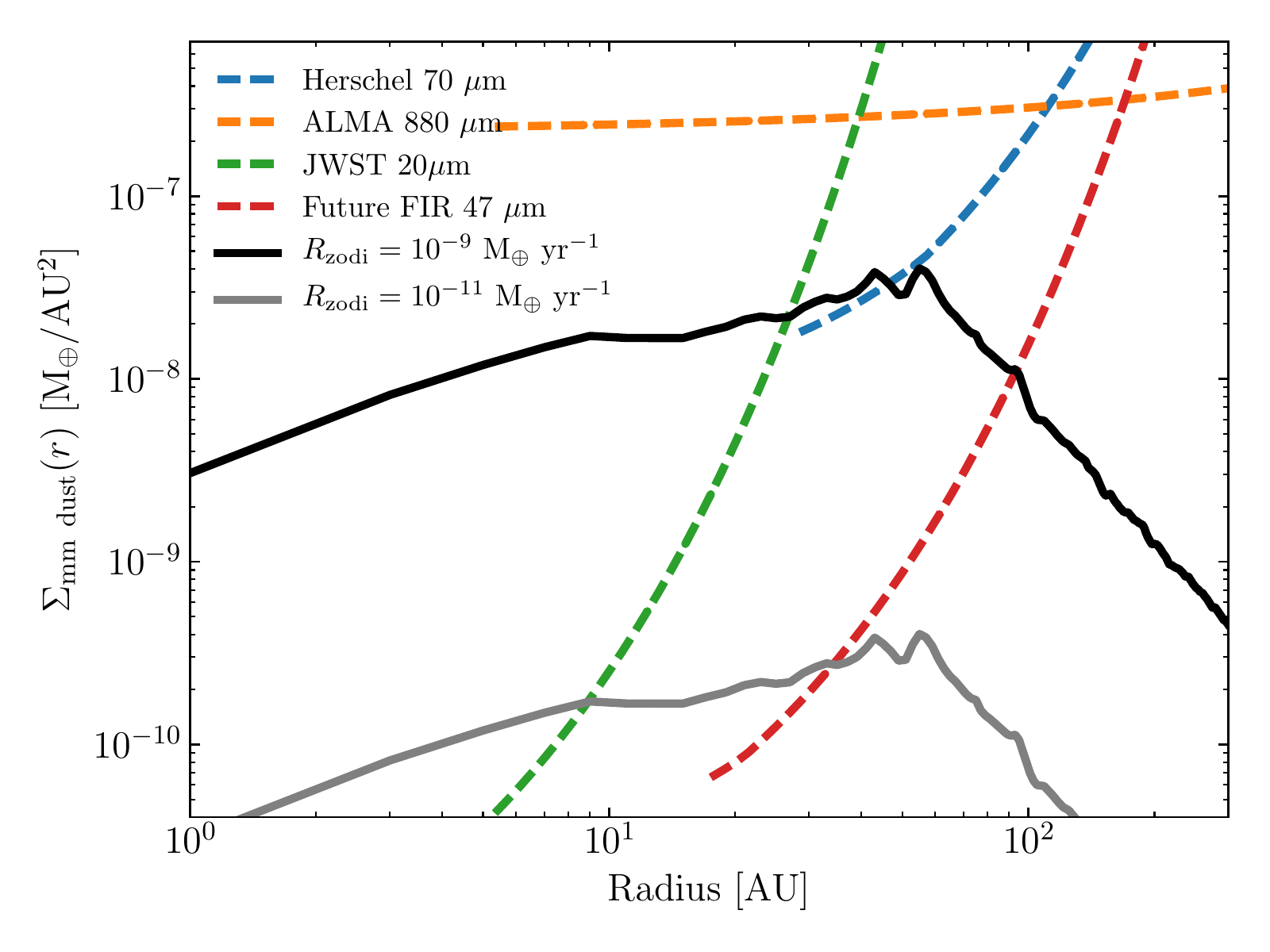}
 \caption{Steady state surface density distribution of dust grains
   smaller than 1 cm for our reference system.  The dashed lines
   represent the 1~h integration sensitivities ($3\sigma$) from
   different instruments, assuming a distance of 10 pc, a face-on disc
   orientation, and azimuthally averaging the emission over 10 AU wide
   annuli. The estimated sensitivities only extend inwards to half of
   the resolution. Note that for the ALMA sensitivity we have not
   considered the size of its primary beam as this can be overcome by
   multiple pointings (mosaic mode).}
 \label{fig:Sr_sen}
\end{figure}

In Figure \ref{fig:Sr_sen} we show the predicted mass surface density
of dust smaller than 1 cm for a system inferred to have
$R_\mathrm{zodi}=10^{-9}$ and $10^{-11}$~\Me~yr$^{-1}$, assuming these
exozodis are fed by our reference chain of planets ($\fin=3.7\%$). The
resulting surface densities are $\sim10^{-8}-10^{-10}$~\Me~AU$^{-2}$
within 50 AU. Figure \ref{fig:Sr_sen} also compares these with typical
sensitivities ($3\sigma$ for 1h observations) of ALMA at 880 $\mu$m
(0.1 mJy and $1\arcsec$ resolution), Herschel at 70 $\mu$m (3.0 mJy
and $5\farcs6$ resolution), JWST at 20 $\mu$m (8.6~$\mu$Jy and
$1\arcsec$ resolution) and a possible future 3-meter far-IR (FIR)
space telescope similar to SPICA at 47~$\mu$m (15~$\mu$Jy and
$3\farcs4$ resolution). We assume a system at a distance of 10 pc with
a 1~$L_\odot$ central star. To translate the above sensitivities to
dust masses, we assume black body temperatures and dust opacities
corresponding to dust grains with a -3.5 size distribution and
composed of a mix of amorphous carbon, astrosilicates and water ice
\citep[e.g.,][]{Marino201761vir}. Moreover, assuming a face-on disc
orientation sensitivities are also improved by azimuthally averaging
the emission over 10~AU ($1\arcsec$) wide disc annuli. Figure
\ref{fig:Sr_sen} shows that although ALMA has one of the highest
resolutions, it is not very sensitive to the dust emission within an
outer belt at $\sim50$~AU. JWST is more sensitive than ALMA within
$\sim30$~AU where dust is expected to be warmer and therefore emits
significantly more at mid-IR wavelengths. In the FIR Herschel does
better than ALMA within $\sim100$~AU and could detect the scattered
disc in systems with extreme exozodis; however, it is limited by its
low resolution. A future FIR mission could do much better, being able
to detect much fainter emission at 47~$\mu$m and resolve structure
down to 20~AU. We conclude that current or previous instruments like
ALMA or Herschel could only constrain the architecture of systems with
the highest mass loss rates (see example below).


\subsubsection{$\eta$~Corvi}
\label{sec:etacorvi}

$\eta$~Corvi is one of the best studied systems with hot/warm dust
\citep[$\sim400$~K,][]{Stencel1991, Smith2009etacorvi}. Its hot
component has a fractional luminosity of $\sim3\times10^{-4}$ and its
location was constrained to be between $\sim0.2-1.4$~AU
\citep{Defrere2015, Kennedy2015lbti, Lebreton2016}. This implies that
it can only be explained if material is resupplied from further out
(e.g. an outer belt of planetesimals) at a rate of
$\sim10^{-9}$~\Me~yr$^{-1}$ \citep{Wyatt2007hotdust}. In fact, this
system is known to host a massive cold debris disc located at around
150~AU and resolved in the sub-millimetre and FIR \citep{Wyatt2005,
  Duchene2014, Marino2017etacorvi}. Despite the presence of this
massive outer belt which could feed an exozodi through small dust
migrating due to P-R drag, this scenario has been discarded as an
explanation for the large exozodi levels observed since it is not
efficient enough \citep{Kennedy2015prdrag}. Moreover, observations
found no dust located between its hot and cold components. However,
\cite{Marino2017etacorvi} did find evidence for CO gas at $\sim20$~AU
using ALMA observations. The short-lived CO gas hints at the
possibility of volatile-rich material being passed inwards from the
outer belt and outgassing, consistent with spectroscopic features of
the hot dust \citep{Lisse2012}. An outstanding question though is
whether this can be achieved without requiring scattered disc
densities that exceed the detection limits of FIR and sub-millimetre
observations.

\begin{figure}
  \centering
 \includegraphics[trim=0.0cm 0.0cm 0.0cm 0.0cm, clip=true, width=1.0\columnwidth]{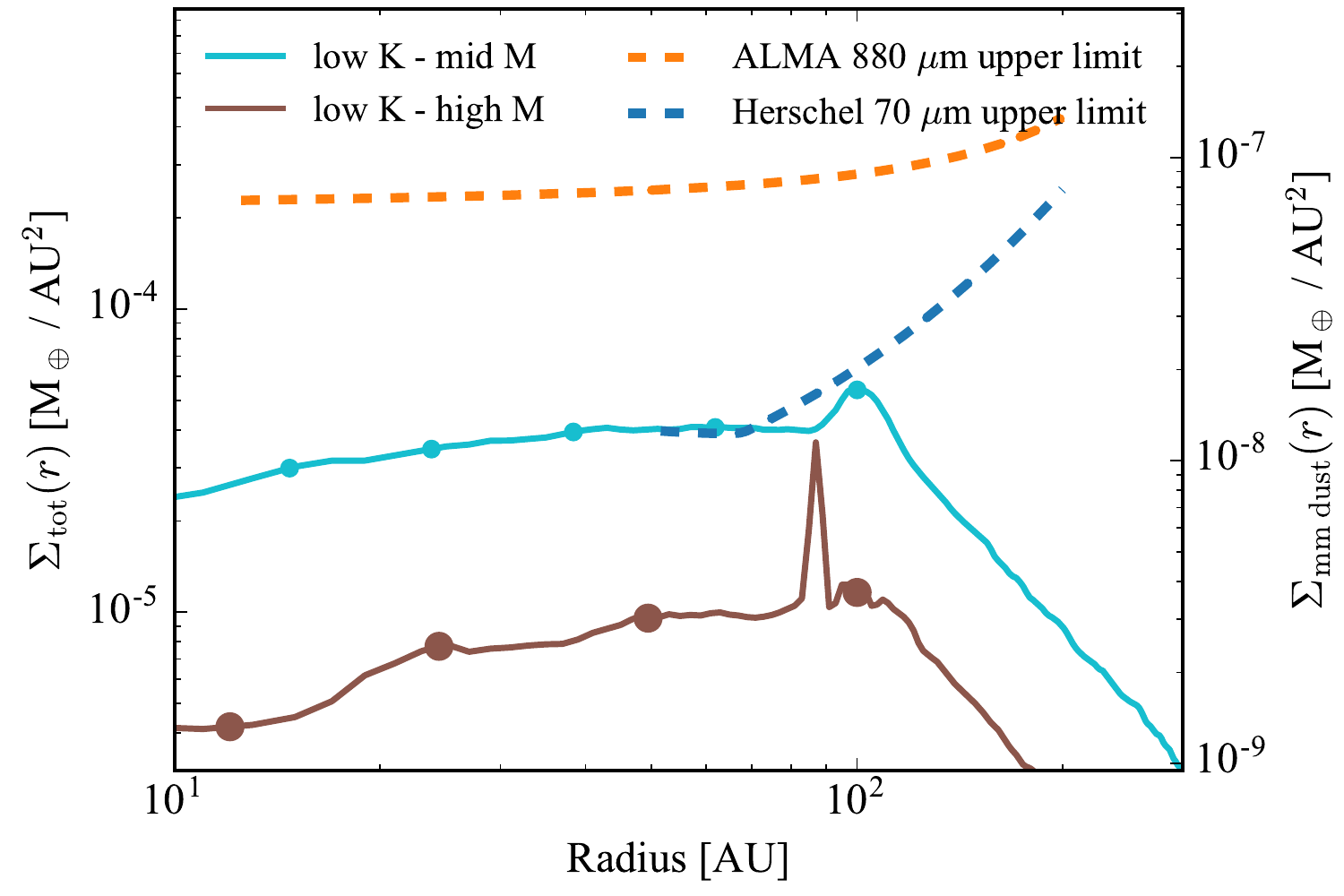}
 \caption{Steady state surface density distribution of particles for a
   model tailored to $\eta$~Corvi. Particles are input in the chaotic
   zone of the outermost planet ($\aplti=100$~AU) at a constant rate
   of $10^{-9}/\fin$~\Me~yr$^{-1}$. All planets are spaced with $K=12$
   and have masses of 30 (light blue) and 90 \Me\ (brown). The dashed
   lines represent the upper limits on the amount of mm-sized dust
   from ALMA at 0.88 mm (blue) and Herschel at 70 $\mu$m (orange).}
 \label{fig:Sretacorvi}
\end{figure}

We run two new models with 30 and 90~\Me\ planets spaced with $K=12$
(tightly packed) to achieve a high inward scattering efficiency, but
now extending the chain of planets up to 100 AU --- the maximum
semi-major axis of a planet sculpting the inner edge of the outer belt
\citep{Marino2017etacorvi}. In the first case, when extending the
chain up to 100~AU we find that $\fin$ decreases from 4.8\% to 2.8\%,
$\tej$ increases from 62 to 120~Myr and $\fej$ increases from 94\% to
97\%. For the more massive planets, we find $\fin=2.1\%$,
$\tej=20$~Myr and $\fej=97\%$. We compute the expected total surface
density for a mass input rate such that
$R_\mathrm{zodi}=10^{-9}$~\Me~yr$^{-1}$. Under the same assumption
stated in the previous section, we extrapolate $\Sigma(r)$ to the
surface density of dust grains smaller than 1 cm
($\Sigma_\mathrm{mm\ dust}$) assuming a -3.5 size distribution between
the largest planetesimals and the cm-sized grains and an opacity of
1.7 cm$^{2}$~g$^{-1}$ at 880~$\mu$m. However, given the existing
constraints from image and spectral energy distribution modelling, we
adopt a grain opacity index $\beta=0.2$ (i.e a $\sim-3.1$ grain size
distribution) between FIR and sub-mm wavelengths. We also assume a
dust temperature of 50~K at 100~AU which increases towards smaller
orbital radii as r$^{-0.5}$, consistent with radiative transfer modelling of
this system. In Figure \ref{fig:Sretacorvi} we compare
$\Sigma_\mathrm{mm\ dust}$ from our simulations with the upper limits
from ALMA and Herschel observations \citep{Marino2017etacorvi,
  Duchene2014}. We find that for both type of systems, the sensitivity
curves of ALMA and Herschel are above the predicted surface
densities. The surface density of a 30~\Me \ chain barely reaches the
3$\sigma$ limit imposed by Herschel. The total mass in dust smaller
than 1 cm in the scattered disc within 80 AU is $3\times10^{-4}$ and
$9\times10^{-5}$~\Me \ for the systems with 30 and 90~\Me\ planets,
respectively. These dust masses are well below the mass upper limit of
$2.7\times10^{-3}$~\Me\ from ALMA observations
\citep{Marino2017etacorvi}.

We can also compare these dust densities with alternative scenarios
such as the P-R drag scenario. Although it is not efficient enough to
explain $\eta$~Corvi's exozodi, P-R drag is inevitable and small dust
will be dragged inwards and be present in between its exozodi and
outer belt with a surface density distribution close to flat. We find
that the two simulated systems presented in this section have optical
depths of $10^{-6}-10^{-5}$, estimated as the product of the surface
density and a standard dust opacity at optical wavelengths (assuming a
-3.5 grain size distribution with a minimum and maximum size of
1~$\mu$m and 1 cm). These optical depths are similar or slightly
greater than for small dust migrating in due to P-R drag in the
absence of planets \citep[see Figure 1
  in][]{Kennedy2015prdrag}. Therefore it is worth noting that
observations looking for an intermediate component within a few AU in
between bright exozodis and outer belts (e.g. using the Large
Binocular Telescope Interferometer) should consider the possibility
that any detected emission could correspond to a scattered disc rather
than P-R dragged dust. Constraints on the radial profile of the
surface density of dust could help to disentangle between these two
scenarios as a scattered disc could have a steeper slope.

\begin{figure}
  \centering
 \includegraphics[trim=0.0cm 0.0cm 0.0cm 0.0cm, clip=true, width=1.0\columnwidth]{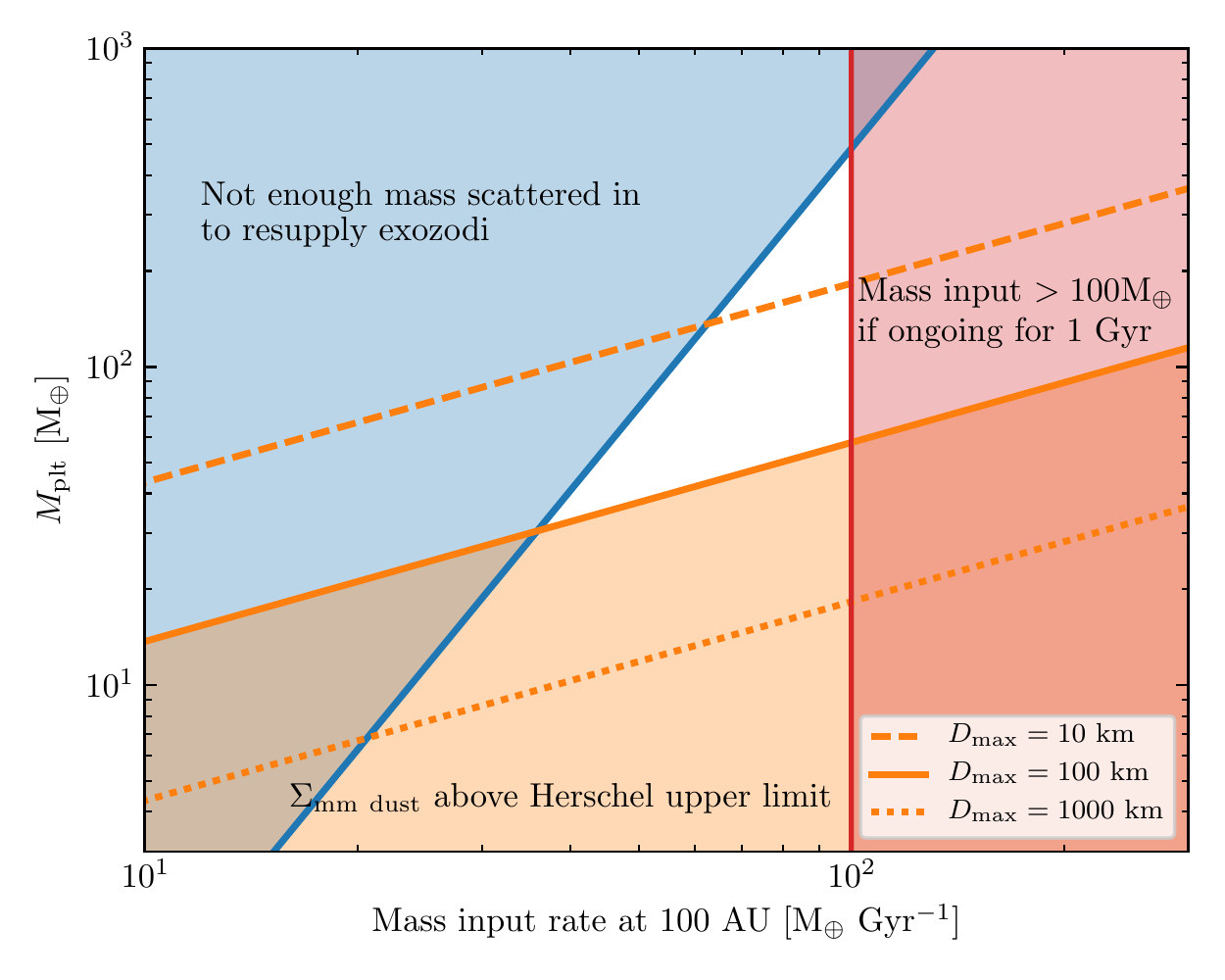}
 \caption{The planetary system that could orbit $\eta$~Corvi and
   supply the exozodi, whilst not contradicting observational limits
   for the scattered disc, and assuming a constant mass input rate at
   100 AU and a system composed of equal mass tightly packed planets
   ($K=12$). The lower limit for the planet mass (orange) is estimated
   based on our result for the surface density as a function of the
   planet masses, the Herschel 70~$\mu$m upper limit for dust between
   $\eta$~Corvi's outer belt and exozodi, and assuming a size
   distribution of solids up to a size of 10, 100 and 1000 km (dashed,
   continuous and dotted orange lines).  The mass upper limit is
   estimated by setting the rate at which mass is transported into 0.5
   AU equal to the exozodi mass loss rate. The red region is discarded
   as it would lead to unphysically large total solid mass beyond 100
   AU being scattered in for 1-2 Gyr, although the exact upper limit
   for the mass input rate is uncertain.  The white region represents
   the planet masses and mass input rate that are roughly consistent
   with the observational constraints and exozodi estimated mass loss
   rate.}
 \label{fig:mplt_etacorvi}
\end{figure}

Now, we can go one step further and use our results to constrain the
planet masses assuming that the exozodi in $\eta$~Corvi is fed by
scattering within a chains of equal mass planets. In \S\ref{sec:varM}
we found that in systems of equal mass planets and uniform spacing (20
mutual Hill radii) the surface density scales with planet mass
approximately as $\Mplt^{-1.6\pm0.2}$, while the fraction of particles
that get to 0.5 AU scales as $\Mplt^{-0.37\pm0.13}$. Assuming these
relations stay the same for planet chains out to 100 AU and with low
spacing (12 mutual Hill radii), we can approximate $\fin$ and the
surface density of dust smaller than 1 cm at 60 AU
($\Sigma_\mathrm{mm, 60~AU}$) by
\begin{small}
    \begin{eqnarray}
      \fin=0.028\left(\frac{M_\mathrm{p}}{30~\mathrm{M}_{\oplus}}\right)^{-0.37}, \label{eq:finetacorvi}\\
      \Sigma_\mathrm{mm, 60~AU}=1.3\times10^{-8} \left(\frac{M_\mathrm{p}}{30~\mathrm{M}_{\oplus}}\right)^{-1.6} \left(\frac{R_\mathrm{in}}{40\ \mathrm{M}_{\oplus}~\mathrm{Gyr}^{-1}}\right) \mathrm{M}_{\oplus}~\mathrm{AU}^{-2}, \label{eq:Sretacorvi}
    \end{eqnarray}
\end{small}
where we have assumed a maximum planetesimal size of 100 km and fixed
the values of $\fin$ and $\Sigma_\mathrm{mm, 60~AU}$ to the ones
presented above, only valid for chains of equal mass 30 \Me\ planets
spaced by 12 mutual Hill radii. By equating Equation
\ref{eq:Sretacorvi} to the Herschel upper limit of $1.3\times10^{-8}
\mathrm{M}_{\oplus}~\mathrm{AU}^{-2}$ at 60 AU we can infer a maximum
planet mass as a function of the mass input rate (orange solid line in
Figure \ref{fig:mplt_etacorvi}). This limit assumes a maximum
planetesimal size ($D_{\max}$) of 100~km and a -3.5 power law size
distribution. A smaller $D_{\max}$ would make $\Sigma_\mathrm{mm,
  60~AU}$ greater (dashed line), while a larger maximum planetesimal
size would decrease the surface density of dust (dotted line). Larger
planet spacing would also have an effect on the surface density, being
increased for larger planet spacing as found in
\S\ref{sec:res_spacing} (e.g. $K=20$).

We also expect that there is a limit on the mass input rate from an
outer belt if it has been ongoing for 1-2 Gyr (the age of
$\eta$~Corvi) at a constant rate. This limit is chosen such that the
total mass scattered from the outer regions is $\lesssim100$~\Me
(close to the solid mass available in a protoplanetary disc at this
location) which is equivalent to a rate of
$\lesssim100$~\Me~Gyr$^{-1}$. This upper limit for the mass input rate
is represented as a red region in Figure \ref{fig:mplt_etacorvi},
although the exact value is uncertain. This upper limit sets a lower
limit for the fraction of scattered particles that get to 0.5 AU of
$\sim$1\% to resupply the exozodi. A caveat in this argument is that
it assumes that mass has been continuously scattered in over
$\sim1$~Gyr at a steady state. However, this could be significantly
different if ongoing only for a fraction of the age of the system or
if stochastic processes are at play, making $R_\mathrm{zodi}$ vary
significantly over short periods of time \citep[e.g. as explored in
  the context of white dwarf pollution,][]{Wyatt2014}.

In addition, because the fraction of particles that get to 0.5 AU
depends on the planet mass (Equation \ref{eq:finetacorvi}), for a
given mass input rate there is a maximum planet mass. This upper limit
is represented as a blue region in Figure \ref{fig:mplt_etacorvi}
assuming chains of equal mass planets spaced by 12 mutual Hill
radii. Planet chains with lower spacing would have a larger $\fin$
therefore the maximum mass could be pushed up, although planet chains
with spacing smaller than 10 mutual Hill radii are likely to go
unstable on Gyr timescales. Wider planet spacing would result in a
lower $\fin$, thus a smaller mass upper limit. It is important to note
that these limits are conservative, as $\fin$ is the maximum fraction
of the material that could be incorporated into the exozodi, since
some of that material that makes it to 0.5 AU will end up being
ejected, which would narrow the allowed white region in Figure
\ref{fig:mplt_etacorvi}.

Based on these excluded regions we can constrain the mass of the
planets scattering material inwards and resupplying the
exozodi. Assuming planets are of equal mass, spaced by 12 mutual Hill
radii and a maximum planetesimal size of 100 km, we find that the mass
input rate and planet mass must be greater than 36~\Me~Gyr$^{-1}$ and
30~\Me, respectively. If the maximum planetesimal size is larger
(smaller), then the planet mass and mass input rate could be smaller
(larger). In addition, assuming a maximum mass input rate of
$\sim100$~\Me~Gyr$^{-1}$, we find a maximum planet mass of
$300$~\Me. Planets much more massive than that are not efficient
enough at scattering particles into 0.5 AU. Future observations by
JWST or FIR space missions could provide important constraints for
the models explored here. A detection of dust between the outer belt
and exozodi together with further modelling considering a wider
variety of planetary system architectures and P-R drag, could be used
to constrain the planet masses in this planet scattering scenario.

\subsection{Best system at transporting material within 0.5 AU}

We have found that the systems best suited for transporting material
inwards are those with lower mass planets and that are tightly packed
(at the limit of stability). Varying the planet spacing and mass as a
function of orbital radius did not result in a higher fraction that
reached 0.5 AU. Moreover, although we varied to extremes the
architecture of these planetary systems to study how the inward
scattering efficiency changes, the fraction that reached 0.5~AU did
not change by more than a factor of $\sim7$, while the surface density
of the scattered disc varied by two orders of magnitude. This could
imply that in order to explain the short-lived hot dust observed in
many systems, no fine tuning is necessary, and a vast range of
architectures could explain this frequent phenomenon.


Because in our analysis we assume that particles are immediately lost
and converted into dust when crossing the inner boundary at 0.5 AU,
our results should be considered as an upper limit or as the maximum
fraction of the mass scattered from the outer regions that could feed
an exozodi assuming all the mass is processed into small dust with a
100\% efficiency. In reality, it is unlikely that all the mass
scattered inside of 0.5 AU will inevitably become the small dust that
is observed as an exozodi. The exact mass fraction transformed into
dust will depend on the specific mechanism that is at play and
detailed modelling is needed. If dust is released by cometary
activity, it would be necessary to consider the outgassing
\citep[e.g.,][]{Marboeuf2016} and disruption process of exocomets,
together with the collisional evolution and radiation forces acting on
dust grains. How fast the mass of a comet can be released in the form
of dust via sublimation or disruption \citep[spontaneous or due to
  tidal forces,][]{Weissman1980, Boehnhardt2004} is uncertain;
however, an order of magnitude estimate can be obtained from
Jupiter-family comets (JFC) that have typical physical lifetimes of
$10^4$~yr \citep{Levison1997}. Hence, exocomets could disintegrate
into small dust on timescales much shorter than the scattering
diffusion timescale for the planet masses that we explored (Equation
\ref{eq:teject} and Figure \ref{fig:setups}). Therefore, removing
particles immediately after they cross our inner boundary could be a
reasonable approximation. Alternatively, the dust could arise from
collisions of scattered planetesimals with an in situ planetesimal
belt, although this may be inefficient if the collision probabilities
are low.




\subsection{Best system at delivering material to inner planets}


We found that the best systems at delivering material to inner planets
were those with low mass outer planets and either with uniform mass or
increasing towards smaller semi-major axis. This is a consequence of
two factors. First, particles being scattered by low mass planets can
stay in the system for longer, increasing the density of particles
near the planets, and so the probability of being accreted by a planet
before being ejected. Second, particles scattered by lower mass
planets tend to have lower eccentricities and inclinations, thus lower
relative velocities which increases the collisional cross-section of
the planets. The latter is important as it also implies that the best
systems at creating exozodis are not necessarily the best at
delivering material to the inner planets (compare green, pink and grey
points in bottom right panel of Figure \ref{fig:results_mass}). For
example, a system with low mass tightly packed planets scatters twice
the amount of particles into 0.5 AU compared to our reference system,
but with a similar fraction of accreted material per inner planet.

Similar to the fraction of material transported into 0.5 AU, the
efficiency of delivery to inner planets does not vary by more than an
order of magnitude when comparing the different systems that we
simulated. We find typical values between 0.1\% to 1\% for the
fraction of scattered particles that are accreted per inner
planet. These could be even higher for lower mass planets (e.g. Earth
or super-Earths) if we extrapolate our results. Note that these
fractions can be understood as collision probabilities, which are
considerably higher than the $10^{-6}$ collision probability of
cometesimals with Earth in the early Solar System
\citep{Morbidelli2000, Levison2000}. It is unclear though how these
results extrapolate to close-in planetary systems of chains of
super-Earths, or close-in Earth-sized planets around low mass stars
\citep[e.g. Trappist-1,][]{Gillon2017}. Most of these planets are in a
different regime where they are much more likely to accrete particles
rather than to eject them \citep[][]{Wyatt2017}, and thus the
accretion efficiencies and relative velocities could be very different
\citep{Kral2018trappist1}.

The low collision probability of comets has been used as an argument
for the unlikely cometary origin for the water on Earth. However, our
results show that this conclusion cannot be simply extrapolated to
extrasolar systems, as the low probabilities might be heavily
determined by the presence of Jupiter and Saturn in the Solar System
which eject most of the minor bodies that are scattered in from the
outer regions never getting into the inner regions. We have shown that
comet delivery can be much more efficient for other architectures and
so could represent a significant source of water and volatiles for
close in planets, although it is not clear yet how common are the
architectures that we assumed in this paper. For example, assuming a
total mass of 1 \Me\ of icy exocomets being scattered with an ice mass
fraction of 0.5 dominated by water \citep[roughly what is found in
  Solar System comets, see review by][]{Mumma2011}, the total amount
of volatiles accreted could be higher than $10^{-4}$~\Me\ per inner
planet, and if extrapolated to systems of 1 \Me \ planets, this could
be enough to deliver the mass of Earth's oceans and atmosphere
($2\times10^{-4}$~\Me) to Earth-like planets. 

An important caveat in our results regarding planet accretion is the
uncertain fraction of volatiles (including water) that a planet is
able to retain from an impact. This strongly depends on the impact
velocities, mass of the planets, volatile fraction of planetesimals,
presence of primordial atmospheres and size of impactors
\citep[e.g.,][]{deNiem2012}. These considerations are beyond the scope
of this paper and require a statistical analysis of the impact
velocities with a larger number of test particles than considered
here.







\subsection{Idealised planetary system}



In this work we have considered idealised systems with regular or
ascending/descending planet masses and spacings as a function of
semi-major axis in order to test how scattering depends on planet
properties. We do not know whether these idealised architectures
really occur in nature, although some degree of regularity could be
common \citep[e.g.][]{Weiss2018, Millholland2017}. Irregularities such
as the presence of a very massive planet \citep[e.g. ejector planet as
  described in][]{Wyatt2017} in between a chain of lower mass planets
(e.g. Jupiter and Saturn in the Solar System) could radically change
the lifetime of particles and the probability of being scattered into
the exozodi regions. Moreover, for certain planetary system
architectures resonant effects may be important for scattering and
could produce results that differ significantly from the trends that
we found here. Therefore further study is necessary to see how our
results can be generalised over a broader range of architectures.

\subsection{Migrating from exo-Kuiper belt to outermost
  planet}
\label{sec:disinwards}

What could drive particles on stable orbits in the outer regions
towards the outer planets? In this paper we have assumed this happens
at a constant rate and for long timescales. We now discuss possible
mechanisms that could make these particles migrate in. For example,
small dust in an outer debris belt could migrate in through P-R drag,
encountering the outermost planet and being scattered further in by
the chain of planets. However, this mechanism cannot produce a high
mass input rate near the outermost planet as it only affects the small
dust. Therefore, a mechanism arising from gravitational interactions
affecting larger bodies is required. Particles could be slowly excited
onto orbits with higher eccentricities and pericentres near the
outermost planet by: chaotic diffusion produced by high order or
three-body resonances acting on long timescales \citep{Duncan1995,
  Nesvorny2001, Morbidelli2005}; secular resonances if multiple
planets are present \citep{Levison1994}; mean-motion resonances with
an exterior massive planet on a low eccentric orbit \citep{Beust1996,
  Faramaz2017}; dwarf planets embedded in the outer belt dynamically
exciting smaller bodies \citep{Munoz-Gutierrez2015}; or Kozai
oscillations induced by an outer companion. Alternatively, if the
outer belt is massive enough the outermost planet could migrate
outwards instead, while scattering material inwards and continually
replenishing its chaotic zone \citep{Bonsor2014}. In these scenarios,
particles that were initially near the outermost planet in the chain
would start interacting with it with similar orbital parameters to
those assumed here, thus we expect that the assumed initial conditions
in this paper are representative of the scenarios stated above. A
fourth possibility is that exocomets are fed from a massive exo-Oort
cloud where they are perturbed by Galactic tides or stellar passages,
decreasing their pericentres enough to have planet-crossing orbits
\citep[e.g.,][]{Veras2013, Wyatt2017}. In this last scenario, we
expect that particles will start interacting with the planets at high
eccentricities and inclinations, and the detailed origin might affect
conclusions that rely on the Tisserand parameter (such as about the
spacing of the outermost planets). All these scenarios could act on
long-timescales feeding material into the vicinity of the outermost
planet in the chain. Although in this paper we have focused purely on
the process of scattering (see \S\ref{sec:framework}), our results are
independent of the rate at which these mechanisms can feed material
into the vicinity of the outermost planet. Understanding how these
different mechanisms can be coupled with the process of inward
scattering by a chain of planets is the subject of future work.





\subsection{Simulation parameters}
\label{sec:simparameters}

Here we discuss the effect of changing some of the chosen simulation
parameters. Throughout this paper we have assumed planets with a
uniform bulk density of 1.6~g~cm$^{-3}$ (Neptune's density), although
planet densities could vary as a function of planet mass as suggested
by Solar System and extrasolar planets \citep[e.g.][and references
  therein]{Chen2017}. For the range of planet masses used in this
paper (4 to 180 \Me), planet densities are expected to vary roughly as
$\sim\Mplt^{-0.8}$ \citep{Chen2017}, thus uniform density assumption
could underestimate and overestimate the planet densities by a factor
of 2.4 for 10 and 90~\Me~planets, respectively. This translates to a
factor of 1.3 in planet radius. For high mass planets, the collisional
cross section is proportional to both mass and radius (due to
gravitational focusing), thus the fraction of accreted particles could
be underestimated by 30\%. On the other hand, the collisional cross
section of low mass planets orbiting within 10~AU could be either
proportional to $\Rplt^2$ (if the relative velocities are greater than
the escape velocity) or $\Rplt\Mplt$ (if gravitational focusing is
important). Hence, the fraction of accreted particles could be
overestimated by 25-45\% for low mass planets. Therefore, the trend
seen in Section \ref{sec:varM} could become flatter if we considered
densities varying as a function of planet mass.

We also tested the effect of changing some other parameters by:
varying the initial distribution of eccentricities and inclinations,
extending the length of our simulations, and varying the inner
boundary to shorter orbital radii. Depending on the specific mechanism
inputting material near the outermost planet, particles will have
different distributions of eccentricities and inclinations. When
varying the initial eccentricity and inclination distributions in our
reference system, we found that $\tej$, $\fin$, $\faccplt$ and
$\Sigma(r)$ do not change significantly, except when the
eccentricities or inclinations are initially very high
(e.g. $e\gtrsim0.2$, see details in appendix
\ref{sec:vareandi}). Higher initial eccentricities led to slightly
higher fractions of particles reaching 0.5 AU and being accreted by
inner planets. This could be due to a larger fraction of particles
having a Tisserand parameter low enough to reach the second outermost
planet.

To test if our results are robust against extending our simulations in
time, we continued the simulation of low mass planets (pink), which is
the one with the slowest evolution, to 5 Gyr. We found that the new
values of $\tej$, $\fin$, $\faccplt$ and $\Sigma(r)$ are consistent
with the ones obtained before (remember that these had been
extrapolated to $t=\infty$ on the assumption that the different
outcomes occurred in the same proportions as they had up to 1 Gyr),
confirming that 1 Gyr of simulation is enough to understand the
behaviour of such systems (see details in appendix \ref{sec:longer}).

Finally, we moved our inner boundary from 0.5 to 0.1 AU finding no
significant changes in $\tej$, $\faccplt$. As expected $\fin$ is
decreased and $\Sigma(r)$ becomes flatter within 4 AU as predicted in
\S\ref{sec:res_reference} (see details in appendix
\ref{sec:inneredge}).





\section{Conclusions}

\label{sec:conclusions}

In this paper we have studied the evolution of exocomets in
exoplanetary systems using a set of N-body simulations. We focused
specifically on how the scattering process varies as a function of the
architecture of the planetary system when exocomets originate in the
outer region of a system (50 AU), e.g. in an outer belt, and start
being scattered in by the outermost planet. We are interested in the
delivery of material to the terrestrial region, either as exozodiacal
dust or cometary material (including volatiles) onto planets
themselves.

This work aims to assess whether exocomets scattered by planets could
provide a plausible explanation for exozodiacal dust commonly observed
in exoplanetary systems. We find that the systems of tightly packed
low mass planets lead to the highest fraction ($\sim7\%$) of scattered
comets reaching the inner regions, where they could resupply
exozodis. Moreover, for a given pair of outermost planets, systems
with decreasing (increasing) planet spacing and mass towards the star
lead to higher (lower) levels of exozodis. We also find that systems
with a very high mass innermost planet (e.g. 150 \Me) are inefficient
at producing exozodis. However, although tightly packed low mass
planets are the most efficient at feeding exozodis, the fraction of
comets scattered within 0.5 AU does not change by more than a factor
of $\sim7$ (1-7\%) when varying the architecture of the planetary
systems that we tested, noting that our simulations were generally
devoid of planets more massive than 0.3~M$_\mathrm{Jup}$. The fact
that this fraction does not change by orders of magnitude suggests
that many different types of planetary architectures could be
efficient at feeding exozodis, possibly explaining the high frequency
of exozodis around nearby stars.

In addition, we characterise the surface density of scattered comets
between the exozodi region and an outer belt of exocomets. The surface
density can be used as a test for the scattering scenario to resupply
exozodis since feeding an exozodi could require the presence of a
scattered disc that is bright enough to be detectable. First, we find
that the surface density radial profile of the scattered disc between
the planets typically increases with distance to the star instead of
being flat as in a purely P-R drag scenario.


Secondly, unlike the fraction of scattered comets into 0.5 AU
($\fin$), the surface density of comets can vary by two orders of
magnitude and is not directly proportional to the fraction scattered
inwards. For example, systems of tightly packed planets have a higher
$\fin$, but a lower surface density compared to a system of planets
with medium spacing. This implies that for a given exozodi, the amount
of scattered material present between the planets can vary depending
on the specific planetary architecture, with systems of low mass
planets and medium spacing having the highest surface density of
material between the planets. Future space missions like JWST or a FIR
space telescope should be able to detect and characterize scattered
discs in thermal emission around nearby systems with exozodis, setting
tight constraints for the comet scattering scenario. For some systems,
current observational limits already allow us to set some
constraints. For example, the current Herschel and ALMA limits on the
dust emission in between $\eta$~Corvi's exozodi ($<1$~AU) and outer
belt (100-200~AU) can be used to rule out some planetary
architectures. For chains of equal mass planets and tightly spaced (12
mutual Hill radii), we find that only planet masses between 30 and 300
\Me\ could feed the exozodi at a high enough rate and hide an
scattered disc below current upper limits, assuming the exozodi levels
have stayed roughly constant and planetesimals/comets have a maximum
size of about 100 km.

Finally, we have studied the delivery of volatiles by exocomets to the
inner planets via impacts. We found that for a variety of
architectures the delivery of material is relatively efficient. For
every thousand comets scattered, between 1-10 are delivered to each
inner planet. This is efficient enough to deliver the mass in Earth's
oceans if $\sim1$~\Me\ of icy exocomets were being scattered, which is
reasonable considering the expected initial mass of exo-Kuiper
belts. Of the planetary architectures explored in this paper, we found
that chains of low mass planets with medium spacing ($\sim20$ mutual
Hill radii) are one of the most efficient at delivering comets to
inner planets. If the spacing is reduced below $\sim20$ mutual Hill
radii, the fraction of particles scattered to the exozodi region
increases, but the number of impacts per inner planet decreases. This
is because particles scattered by tightly spaced planets evolve faster
and are lost before they can be accreted by a planet. This results in
a lower surface density of the scattered disc for these systems. The
systems that lead to the most planetary impacts have low mass outer
planets and high mass inner planets. This configuration maximises the
collisional cross-section of inner planets as they have high masses
and the particles scattered in by low mass planets have lower relative
velocities. Hence, low mass outer planets are best suited for
delivering material to the inner planets. Our results show that
exoplanetary systems could potentially deliver volatiles to inner
planets at a similar level to Earth, and if chains of low mass planets
are common, they may not lack the volatiles necessary to sustain life.

\section*{Acknowledgements}

MCW and QK acknowledge funding from STFC via the Institute of
Astronomy, Cambridge Consolidated Grant. AB acknowledges Royal Society
via a Dorothy Hodgkin Fellowship.






\bibliographystyle{mnras}
\bibliography{SM_pformation} 



\appendix

\section{Varying the initial \lowercase{$e$} and $I$}
\label{sec:vareandi}

In order to test the effect on our initial conditions on $e$ and $I$,
i.e. $e_{\max}=0.02$ and $I_{\max}=10\degr$, we repeat our simulations
twice for the reference planet configuration, but changing i)
$e_{\max}$ to 0.2 and ii) $I_{\max}$ to $1.1\degr$. In the first case,
the minimum Tisserand parameter and pericentre are reduced from 2.97
to 2.93 and from 18 to 13 AU, respectively. The fraction of particles
ejected stays roughly the same around 93\%, the timescale for ejection
decreases from 120 to 95 Myr, the number of trojans decreases from
30\% to 22\%, $\fin$ increases marginally from $3.7\pm0.8\%$ to
$4.8\pm0.8\%$ and $\faccplt$ decreases from $0.35\pm0.17\%$ to
$0.25\pm0.13\%$, although the changes in $\fin$ and $\faccplt$ are
still consistent within errors. On the other hand, $\Sigma(r)$ stays
roughly constant.

In the second case, the minimum Tisserand parameter increases to 2.999
and the minimum pericentre to 27 AU, near the orbit of the second
outermost planet. The fraction of particles ejected decreases slightly
to 91\%, the ejection timescale increases to 124 Myr, the fraction of
trojans stays the same, $\fin$ increases marginally to $4.1\pm0.9\%$
and $\faccplt$ decreases to $0.15^{+0.16}_{-0.08}\%$, although both
consistent within errors with our reference system. We also find that
$\Sigma(r)$ stays roughly constant.

These results show that reducing the initial inclination of particles
has no significant effect on our results. Increasing the initial
eccentricities, however, could increase slightly the amount of
material that gets to 0.5 AU and decrease the amount of material that
is accreted by inner planets. Note that an eccentricity of 0.2 is at
the limit of what we would expect in a cold exo-Kuiper belt that has
not been perturbed by an eccentric planet
\citep[e.g.,][]{Kenyon2008}. Therefore, we expect that the results and
trends found in this paper are robust against different initial
eccentricities or inclinations.

\section{Varying the length of simulations}
\label{sec:longer}

To test if our results are dependant on the length of our simulations,
we extended the integration to 5 Gyr for the planet configuration that
had the slowest evolution. This is the system of 10 \Me\ planets
spaced with $K=20$. We find that the timescale for ejection was
correctly estimated being $570\pm35$ Myr, even though a significant
fraction of the particles had not been ejected after 1 Gyr. The
fraction of ejected particles increased slightly from 90 to 92\%, but
consistent within errors. We also find that $\fin$ and the fraction of
accreted particles per inner planet is slightly lower, but consistent
with our previous estimate given uncertainties. Finally, we find that
the derived steady state surface density is consistent with one
derived only considering 1 Gyr of evolution. We conclude that 1 Gyr is
enough time to understand the behaviour of these systems with outer
planets with masses $\gtrsim10$~\Me~and our results and conclusions
would not change significantly by extending the length of our
simulations.


\section{Varying the inner boundary to 0.1 AU}
\label{sec:inneredge}

In our simulations we remove particles with a pericentre lower than
0.5 AU for two reasons, to keep track of how many particles get to the
very inner regions, and because we cannot rely on orbits within 0.5 AU
as we use a time-step of 30 days. To explore the effects that this
causes for example in the surface density and the number of impacts on
inner planets, we moved our inner edge to 0.1~AU, reducing the
time-step of the integration to 3 days, but without adding extra
planets. As expected the number of particles that cross the inner edge
decreased from 3.7\% to 1.5\%. The fraction of particles ejected
increased slightly and the fraction accreted stayed roughly
constant. We also find that the surface density only changes within 4
AU, where it is higher compared to our reference system as particles
that were previously removed stay in the system for longer. The
fraction of particles accreted per inner planet did not increase
significantly because only one of the three inner planets resides
within 4 AU, where $\Sigma(r)$ increased.  Moreover, the new particles
able to remain with the lower inner edge are highly eccentric, hence
less likely to be accreted. Based on this surface density, we derive a
power law index of 0.3 between 1 and 50 AU, flatter than derived
before.




\bsp	
\label{lastpage}
\end{document}